\documentclass[%
 reprint,
superscriptaddress,
 amsmath,amssymb,
 aps,
 prx,,
]{revtex4-2}

\usepackage{graphicx}
\usepackage[english]{babel}
\usepackage{svg}
\usepackage{xcolor}
\graphicspath{{Figures/}}
\usepackage{dcolumn}
\usepackage{bm} 
\usepackage{physics}
\usepackage{booktabs}
\usepackage{subcaption}
\usepackage{cprotect}
\usepackage{hyperref}
\hypersetup{colorlinks=true,allcolors=blue}
\usepackage{float}
\usepackage[capitalise]{cleveref}
\usepackage{tabularx}
\usepackage{url}

\begin{document}


\title{A feature-based information-theoretic approach for detecting interpretable, long-timescale pairwise interactions from time series}

\author{Aria Nguyen}
\affiliation{School of Physics, The University of Sydney, Camperdown NSW 2006, Australia} 
\affiliation{Centre for Complex Systems, The University of Sydney, Camperdown NSW 2006, Australia} 
\author{Oscar McMullin}
\affiliation{School of Physics, The University of Sydney, Camperdown NSW 2006, Australia} 
\author{Joseph T. Lizier}
\affiliation{Centre for Complex Systems, The University of Sydney, Camperdown NSW 2006, Australia} 
\affiliation{School of Computer Science, The University of Sydney, Camperdown NSW 2006, Australia} 
\author{Ben D. Fulcher}
\affiliation{School of Physics, The University of Sydney, Camperdown NSW 2006, Australia}
\affiliation{Centre for Complex Systems, The University of Sydney, Camperdown NSW 2006, Australia}

\date{\today}

\begin{abstract}
Quantifying relationships between components of a complex system is critical to understanding the rich network of interactions that characterize the behavior of the system.
Traditional methods for detecting pairwise dependence of time series, such as Pearson correlation, Granger causality, and mutual information, are computed directly in the space of measured time-series values.
But for systems in which interactions are mediated by statistical properties of the time series (`time-series features') over longer timescales, this approach can fail to capture the underlying dependence from limited and noisy time-series data, and can be challenging to interpret.
Addressing these issues, here we introduce an information-theoretic method for detecting dependence between time series mediated by time-series features that provides interpretable insights into the nature of the interactions.
Our method extracts a candidate set of time-series features from sliding windows of the source time series and assesses their role in mediating a relationship to values of the target process.
Across simulations of three different generative processes, we demonstrate that our feature-based approach can outperform a traditional inference approach based on raw time-series values, especially in challenging scenarios characterized by short time-series lengths, high noise levels, and long interaction timescales.
Our work introduces a new tool for inferring and interpreting feature-mediated interactions from time-series data, contributing to the broader landscape of quantitative analysis in complex systems research, with potential applications in various domains including but not limited to neuroscience, finance, climate science, and engineering.
\end{abstract}


\maketitle


\section{Introduction}
\label{sec:introduction}

The world we live in is an immensely complex system woven by myriad interacting processes.
Unravelling this intricate web presents a significant and ongoing challenge in science---it requires deciphering complex interdependence to reveal the underlying mechanisms that drive a system's behaviors, often from incomplete and noisy data \cite{Peel2022StatisticalInferenceNetworkScience}.
Central to this pursuit is the task of identifying pairwise interaction between two processes by inferring statistical dependence from their time-series data.
This step is crucial for characterizing a system and can serve as the building block for more complex problems such as identifying causal relationships, predicting system behaviors, discovering modular structures within the system, and controlling and manipulating complex systems.
Examples of these problems can be seen in various domains, such as causal inference in Earth system sciences \cite{Runge2019CausalInferenceEarthScience}, community detection in social network science \cite{Hoffmann2020ComunityDetection}, and motion prediction and risk assessment for intelligent vehicles \cite{Lefevre2014MotionPrediction}.

A diverse array of pairwise dependence detection methods have been developed, from simple contemporaneous linear dependence, captured using a Pearson correlation coefficient \cite{freedman2007statistics}, to more complex formulations such as Granger causality \cite{Granger2001causality}.
A recent survey implemented and compared over 250 pairwise dependence measures including measures derived from information theory, distance similarity, convergent cross mapping, and others \cite{Cliff2023UnifyingPairwiseInteractions}.
Among these measures, information-theoretic measures such as mutual information \cite{MacKay2003informationtheory} and transfer entropy \cite{Schreiber2000TE,Bossomaier2016TEIntro} are powerful tools for detecting pairwise dependencies in time series in a model-free way, marked by an absence of assumptions about the underlying mathematical or statistical models governing the variables.
These methods quantify the amount of shared information between two processes (or, in other words, the reduction in uncertainty of one process given knowledge of another process) from the raw time-series data and evaluate their statistical dependence.
For example, mutual information (MI) calculates the reduction of uncertainty (or entropy) in one variable given the knowledge of another variable, based on their marginal and joint probabilities estimated from time-series observations \cite{MacKay2003informationtheory}.
Transfer entropy (TE) quantifies the directed information transferred from one process to another by computing the mutual information between the processes given the knowledge of some historical states of the target process \cite{Schreiber2000TE,Bossomaier2016TEIntro}.
These methods can capture both linear and nonlinear dependence from time-series data and have been applied across numerous fields, including fluid mechanics \cite{Materassi2014TEFluidTurbulence}, neuroscience \cite{Ursino2020TEBrainConnectivity}, and finance \cite{Dimpfl2013TEInformationFlowFinancec,Li2013TERiskContagion}.

Despite the popularity of these model-free inference methods, they face significant challenges when dealing with real-world complex systems with long memory, where interactions are mediated by statistical properties of the dynamics computed over extended timescales.
Examples of systems driven by statistical properties of other systems are common in various domains: in financial markets, the volatility of a stock index can trigger cascading effects on others \cite{Li2015StockVolatilitySpillover}, while in neuroscience, communication between brain regions is thought to occurs through the frequencies of neural oscillations \cite{Bonnefond2017OscillationBasedBrainCommunication}.
The typical statistical framing of interactions between time-varying processes, in terms of a small number of recent time-series values themselves, is appropriate for capturing relevant interactions between short-memory processes.
However, in systems with elements that possess long-term memory and computational abilities, there is an advantage in succinctly capturing information and inferring dynamics based on more complex properties of other evolving processes.
This may be particularly relevant to understanding systems in which an observed process exhibits non-stationary variations linked to an underlying variable of interest, such as a time-varying system parameter or non-autonomous drive.
For instance, neural population activity patterns may vary with slow modulation of the excitation-inhibition balance of neural circuits \cite{Markicevic2020EIBrainDisorders, Trakoshis2020EI}, or speech phonemes may vary with a human subject's anxiety level \cite{Albuquerque2021SpeechAnxiety}.
In such cases, responding to the variable of interest through subtle statistical properties of a longer time-series window can be advantageous.
We thus anticipate many adaptive systems with memory and information-processing abilities to interact in ways that exploit non-stationary variations in dynamical properties of the observed processes.

Classic information-theoretic measures like MI and TE, while suitable for low-order Markov processes, can be inefficient in detecting such long-timescale interactions, particularly when dealing with short and noisy time series.
This is because, in remaining model-free, these methods use historical or delayed samples of the variables (referred to as samples from the `signal space') to construct an embedding space representing their underlying state (akin to a Takens embedding \cite{Taken1981TakenEmbeddings}, see \cite[Sec 4.1.1]{Lizier2012Thesis}), and estimate the joint and marginal probability distributions of these embeddings.
For interactions spanning longer timescales and involving numerous historical samples, this means estimating probability densities in a high-dimensional space, which requires a large number of data points that increases exponentially with the dimensionality of the space.
Noisy signals pose a further challenge for capturing the underlying relationships between variables \cite{Edinburgh2021Causality}, a problem that is exacerbated in high-dimensional spaces.
As the available data becomes increasingly sparse in such spaces, distinguishing true signal from noise becomes more difficult: a manifestation of the phenomenon known as the `curse of dimensionality' \cite{Bellman1966DynamicPrograming,Donoho2000HighDimensionDataAnalysis}.
Additionally, the timescale on which processes influence each other is typically unknown, and it is particularly computationally demanding to search for interactions across all possible combinations of historical windows.

Some research has aimed to address the challenges of detecting long-timescale interactions within a information-theoretic framework.
A key example is symbolic transfer entropy, a combination of permutation entropy and transfer entropy, which captures information in the ordinal relationships in the time series, and was demonstrated to be fast and robust to noise \cite{Staniek2008SymbolicTE, Zanin2012PermunationEntropyReview}.
Others have been more targeted to specific forms of interactions, such as the development of a new transfer entropy estimator that is more sensitive to both short- and long-timescale effects in neural spike-train data \cite{Shorten2021TESpiketrain}.
While these adaptations may be efficient at capturing longer timescale interactions, it is crucial to note that they often come with strict assumptions and/or require intimate knowledge of the systems.
When one has limited knowledge of a system, this raises the need for a method capable of detecting long-timescale interactions driven by properties of the dynamics while making no strict assumptions on the systems, and ideally offering interpretability that can provide insights into the nature of interactions.

A potential solution lies in capturing summary statistics of the dynamical properties that may mediate an interaction, allowing dynamics on longer timescales to be meaningfully reduced to interpretable statistical properties.
These summary statistics, or `time-series features', can encompass various statistical properties, ranging from distributional characteristics (e.g., mean, standard deviation, Gaussianity), to dynamical structure (e.g., autocorrelation, stationarity, and periodicity) \cite{Fulcher2018:FeaturebasedTimeseriesAnalysis}.
The temporal variation of these features can then be used as the basis for inferring dependency between two processes, instead of the raw time-series values themselves.
In this way, we may directly assess the statistical dependence between two processes mediated by candidate statistical properties of the processes.
This approach allows for the selection of domain-specific features for an efficient representation of available data.
In situations where domain knowledge is limited, a broad range of generic time-series features can identify interactions arising from different dynamical properties.
Notably, recent years have seen an expansion in such general-purpose time-series feature sets \cite{Trent2021FeatureSetEvaluation}, including: the comprehensive \textit{hctsa} set of over 7000 features \cite{Fulcher2013hctsa, Fulcher2017:HctsaComputationalFramework}; a high-performing subset of 22 features \textit{catch22} \cite{Lubba2019Catch22}; the Python-based \textit{tsfresh} \cite{tsfresh}; and the R-based \textit{feasts} \cite{feasts_pkg}.
These feature sets offer a promising ground for implementing a data-driven feature-based approach for inferring and understanding pairwise dependencies from time-series data.

In this work, we introduce a feature-based adaptation of conventional information-theoretic dependence detection methods that combine their data-driven flexibility with the strengths of time-series features.
The methodology involves first transforming segments of a time series into concise and interpretable summary statistics from a candidate feature set (we will refer to these summary statistics on time-series segments as `windowed time-series features'.)
Mutual information is then used to assess the pairwise dependence between the windowed time-series feature values of the source process and the time-series values of the target process.
This allows us to detect dependence between a pair of time series when the interaction is mediated by a specific statistical feature of the dynamics.
Although this approach involves a trade-off in terms of information and flexibility compared to traditional methods that operate in the signal space, it leverages more efficient representations of the joint probability of source and target processes, which is particularly beneficial for addressing challenges related to high-dimensional density estimation in long-timescale interactions.
Additionally, the interpretability of the features and their grounding in existing scientific theory provides valuable insights into the nature of interactions.
We validate our method on simulated coupled systems with three generative processes: random noise, and two non-stationary processes that capture key characteristics commonly found in real-world processes.
The robustness of the feature-based approach was assessed against the conventional approach across variations in the timescale of interaction, time-series length, and noise level.
We find that the feature-based approach has similar or better performance than the conventional approach when using features that are sensitive the true driving causes of the target processes, indicating promising potential of the method to capture interpretable time-series dependence with a broad feature set.

\section{Methods}
\label{sec:methods}

In this section we introduce a method for detecting a statistical dependence between two processes from measured time-series data.
Specifically, we consider scenarios where the time-series values of one process (`target process') depend on a statistical property of the dynamics across some time window of the other process ( `source process').
Our proposed method uses Mutual Information (MI) between each feature from a set of windowed time-series features of the source and the target time series, which we denote here as $\mathrm{MI}_f$, as detailed in Sec.~\ref{subsec:FIPI}.
We contrast $\mathrm{MI}_f$ with a conventional formulation of MI in this setting, which involves estimating MI directly on the signal space of the time series, denoted here as $\mathrm{MI}_s$.
Using a set $F$ of candidate features, the inference for the dependence between the source and target time series is based on the statistical significance of $\mathrm{MI}_f$ for each $f \in F$.
This feature-based inference of dependence from the set of candidate features $F$ is referred to as $\mathrm{MI}_F$.
Section~\ref{subsec:simulation_studies} outlines the design of our simulation studies for validating and assessing the performance of $\mathrm{MI}_F$ relative to $\mathrm{MI}_s$.
Lastly, we provide an in-depth examination of the generative processes used in our simulation experiments in Sec.~\ref{subsec:generative_processes}.

\begin{figure*}[ht!]
    \centering
  \includegraphics[width=0.85\textwidth]{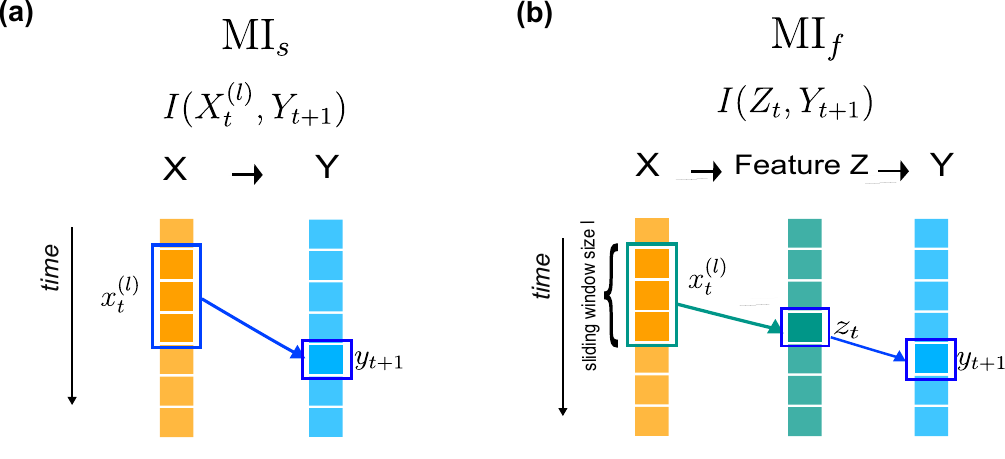}
  \caption{
  \textbf{We introduce a feature-based formulation of Mutual Information, denoted $\bold{MI_f}$, for detecting pairwise interactions between time series.}
  This figure illustrates our method for detecting cases in which a target process, $Y$, is influenced by a statistical property of a recent time window of a source process, $X$.
  We contrast it to conventional MI, estimated directly from the signal-space of the variables, which we denote as $\mathrm{MI}_s$.
  \textbf{(a)} $\mathrm{MI}_s$ is computed based on the observed time-series values of process $X$ and $Y$ [see Eq.~\eqref{eqn:MI_s}].
  \textbf{(b)} $\mathrm{MI}_f$ iterates through time-series segments of length $l$ of process $X$ and reduces each window to a single real-valued summary statistic $z_t$ [see \cref{eqn:feature}].
  MI is then computed between feature variable $Z_t$ and the target variable $Y_{t+1}$ [cf. Eq.~\eqref{eqn:MI_f}].
  We can iterate over a set of candidate features for mapping $\mathbf{x}_t^{(l)}$ to $z_t$, allowing us to detect for interaction between $X$ and $Y$ when it is mediated by a candidate time-series property.
  }
  \label{fig:methodschematic}
\end{figure*}

\subsection{Inferring interactions mediated by windowed time-series features}
\label{subsec:FIPI}

Here we outline our proposed feature-based MI and highlight how it differs from conventional signal-based MI.
The distinction between the two approaches is illustrated schematically in Fig.~\ref{fig:methodschematic}.
Consider two processes, $X$ and $Y$, for which a finite number of samples, $T$, have been recorded over time and represented as the time-series variables $X_t$ and $Y_t$, for $t = 1,2,.., T$.
We aim to determine whether $Y$ is statistically dependent on $X$, and thus refer to $X$ as the `source process' and $Y$ as the `target process'.
Our focus here is to infer a statistical dependence from a specific realization of the source process as a time series, denoted as $\mathbf{x}$, and the target time series, $\mathbf{y}$.

First, we define MI between processes $X$ and $Y$, $I(X;Y)$, based on the probability distribution of their values, as
\begin{equation}
\label{eqn:MI}
I(X;Y) = \sum p(x,y)\log \frac{p(x,y)}{p(x)p(y)}\,,
\end{equation}
where $p(x)$ is the probability that $X$ takes value $x$, $p(y)$ is the probability that $Y$ takes value $y$, $p(x,y)$ is the joint probability that $X = x$ and $Y = y$, and the sum is taken over all possible values of $x$ and $y$ \cite{CoverThomas2005InformationTheory}.

The dynamics of $X$ and $Y$ can be incorporated by using probabilities of consecutive time-series samples, for example, to quantify the information that the history of the source process $X$ provides about the target process $Y$'s next time-series value $Y_{t+1}$ \cite{Schreiber2000TE}.
The past of the source process is contained in its time-delay embedding state $\mathbf{X}_t^{(l)} = (X_{t-l+1},...,X_{t-1},X_{t})$ (as a Takens embedding \cite{Taken1981TakenEmbeddings} with embedding dimension $l$ and time-delay $\tau = 1$) with realization $\mathbf{x}_t^{(l)} = (x_{t-l+1},...,x_{t-1},x_{t})$, where $t = 1,2,..., T$.
The expected MI \cite{Lizier2014JIDT} from the source time-delay embedding state, $\mathbf{x}_t^{(l)}$, to the next target sample, $y_{t+1}$, denoted here as $\mathrm{MI}_s$ to signify its computation on the signal space of the source process, is
\begin{equation}
\label{eqn:MI_s}
 \mathrm{MI}_s = I(\mathbf{X}_t^{(l)};Y_{t+1}) = \sum p(y_{t+1},\mathbf{x}_t^{(l)} ) \log \frac{p(y_{t+1},\mathbf{x}^{(l)}_t)}{p(y_{t+1})p(\mathbf{x}_t^{(l)})} \,.
\end{equation}  
The estimation of MI between a recent window of $l$ values of the source time series and the next value of the target time series is illustrated in Fig.~\ref{fig:methodschematic}(a).

When investigating the response of the target process, $Y$, to longer timescales of the source (i.e., when $l$ is high), estimating the source--target interaction using Eq.~\eqref{eqn:MI_s} requires estimating probability densities in a high-dimensional space, which applies to both the joint probability $p(y_{t+1},\mathbf{x}_t^{(l)})$ and marginal probability $p(\mathbf{x}_t^{(l)})$.
Reliably estimating these probability densities is challenging, requiring a very high number of samples, particularly in the presence of noise.
As shown in Fig.~\ref{fig:methodschematic}(b), $\mathrm{MI}_f$ aims to test whether the target, $Y$, responds to the temporal changes in statistical property of a recent time-window of source, $X$.
This is done by constructing a time-series process $Z$, where the windowed time-series feature $Z_t$, with realization $z_t$, captures a statistical property of the time-series values in the time-delay embedding state $\mathbf{X}_t^{(l)}$ with realization $\mathbf{x}_t^{(l)}$:
\begin{equation}
    z_t = f(\mathbf{x}_t^{(l)}) \,,
\end{equation}
where $f$ is a feature mapping function that takes a time-series segment as input to output a real value that reflects a statistical property of the time-series segment $\mathbf{x}_t^{(l)}$.
For example, one simple choice for a feature-mapping function could be the mean, which takes a time-series segment of $l$ values $x_{t-l+1}, x_{t-l+2},...,x_t$ as input and outputs the average value of these values as $z_t$.
Then, instead of measuring MI between $\mathbf{X}_t^{(l)}$ and $Y_{t+1}$, we compute MI between $Z_t$ and $Y_{t+1}$, denoted as $\mathrm{MI}_f$, as:
\begin{equation}
\label{eqn:MI_f}
\mathrm{MI}_f = I(Z_t;Y_{t+1}) = \sum p(y_{t+1},z_t) \log \frac{p(y_{t+1},z_t)}{p(y_{t+1})p(z_t)} \,.
\end{equation}
We refer to $Z$ as a `capturing feature', as it aims to capture the interaction between $X$ and $Y$, while $l$ is referred to here as the embedding size or `capturing timescale'.
If the values of $Y$ are indeed directly coupled to $Z$, and thus $Z$ captures all of the information in $X$ about $Y$, then by the data processing inequality \cite{CoverThomas2005InformationTheory} we will have $I(Z_t;Y_{t+1}) = I(\mathbf{X}_t^{(l)};Y_{t+1})$.

This procedure can be performed for different statistical properties of $\mathbf{X}_t^{(l)}$ by iterating over local segments of length $l$ of $\mathbf{x}$ with an increment step of 1 sample to create a time-series realization $\mathbf{z}_j$, as
\begin{equation}
\label{eqn:feature}
{(z_j)}_t = f_j(x_{t-l+1}, x_{t-l+2},...,x_t)\,,
\end{equation}
where $f_j$ is the $j$-th feature function from the feature set, $F = \{f_1,f_2,...,f_n\}$, of candidate feature-mapping functions.
Each feature $f_j \in F$ maps the source embedding space, $\mathbb{R}^{l\times {(T-l+1)}}$, to the feature space, $\mathbb{R}^{1\times {(T-l+1)}}$, where $(T-l+1)$ is the number of samples for embedding state $\mathbf{x}_t^{(l)}$, or the number of time-series values $z_t$, where $t = l,l+1,...,T-1$. 
This reduces the dimensionality of the inference from $l$ to 1.
In the case that multiple timescales are hypothesized to mediate the interaction, the process can be repeated across the set of candidate timescales.

Once we have computed $\mathrm{MI}_f$ for each feature $f \in F$, we can proceed to estimate its statistical significance to infer dependence between $\mathbf{X}_t^{(l)}$ and $Y_{t+1}$ \cite{Verdes2005CausalityAssessment,Vicente2011TE,Lizier2014JIDT}.
The likelihood of the observed data under the null model for dependence is derived by generating surrogates of $Z_t$ to estimate the $p$-value of the measures, and then corrected for multiple comparison across the $n$ candidate features using the Holm--Bonferroni method \cite{Holm1979MultipleTest}.
In the case that any feature has a $p$-value less than a given threshold (corrected for multiple comparisons), we infer a dependence between $\mathbf{X}_t^{(l)}$ and $Y_{t+1}$. 
This method, utilizing a set of candidate features $F$ to compute MI measures between the feature time series and the target time series and infer dependence, is denoted as $\mathrm{MI}_F$.

While we have demonstrated the approach here for the case of MI and studied  $\mathrm{MI}_F$ in this work, it is important to note that a similar feature-based modification could also be made to other information-theoretic measures.
For example, Transfer Entropy (TE) \cite{Schreiber2000TE,Bossomaier2016TEIntro} aims to quantify the information that the $l$ historical states of the source process $X$ provides about the target process $Y$'s next time-series value in the context of the target's past, captured by its time-delay embedding $y_t^{(k)} = (y_{t-k+1},...,y_{t-1},y_{t})$, where $k$ is the embedding dimension of the target time series.
We could similarly formulate a feature-based TE by again substituting $Z_t$ for $\mathbf{X}_t^{(l)}$, as
\begin{equation}
\label{eqn:TE_f}
T_{X\rightarrow Y}(l,k) = I(Z_t;Y_{t+1}|Y^{(k)}_t)\,.
\end{equation}

\begin{figure*}[ht!]
    \centering
  \includegraphics[width=1\textwidth]{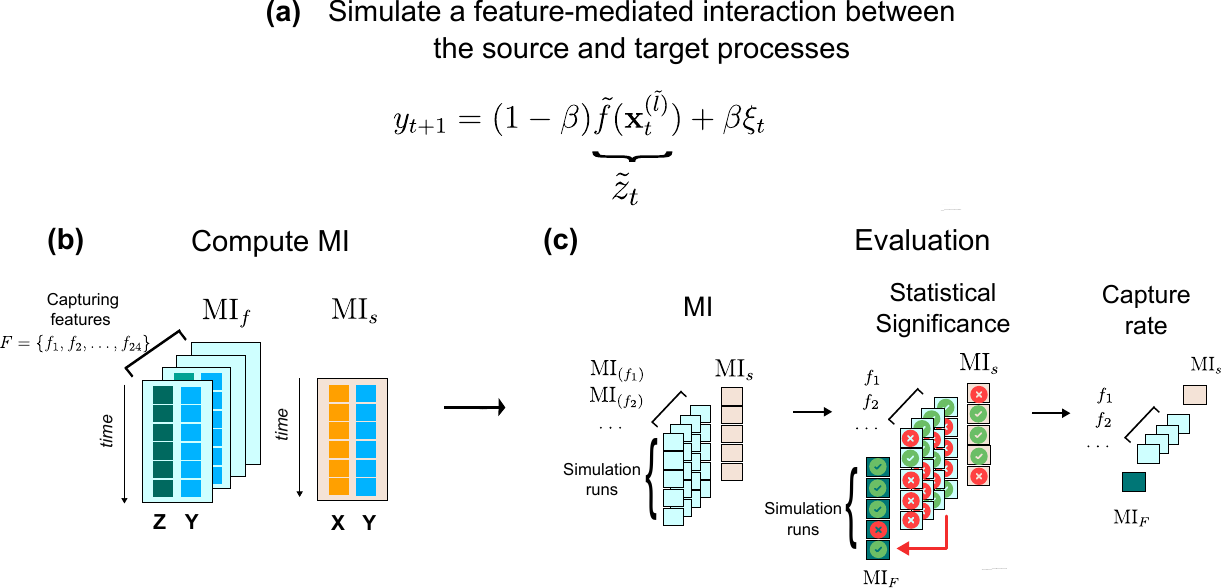}
  \caption{
  \textbf{We simulated pairs of dynamical processes with feature-mediated interactions to evaluate the performance of pairwise dependence methods.}
    \textbf{(a)} We simulated a source and target time series, $\mathbf{x}$ and $\mathbf{y}$ respectively, with a ground-truth feature-mediated dependency.
  Next we computed a feature time series, $\mathbf{\tilde{z}}$, by extracting features across a set of sliding windows of $\mathbf{x}$ (window size $\tilde{l}$) (cf. \cref{eqn:driving_feature}) (note our convention here to use the tilde symbol for ground-truth: $\mathbf{\tilde{z}}$ is the ground-truth feature that drives the interaction between $X$ and $Y$, and $\tilde{l}$ is the ground-truth timescale of interaction).
  Then the target time series $\mathbf{y}$ was generated using a linear function of this feature after standardizing it (cf. \cref{eqn:driving_target}).
  \textbf{(b)} For a given timescale $l$ (from the set of timescales of interest), we computed feature-based MI, $\mathrm{MI}_f$, using a set of 24 capturing time-series features $F$, computed over sliding windows size $l$ that yield a feature time-series $\mathbf{z}$ for each feature in the capturing feature set (cf. Eq.~\eqref{eqn:MI_f}).
  For comparison, we computed $\mathrm{MI}_s$ between $\mathbf{x}_t^{(l)}$ and $y_{t+1}$ (cf. \cref{eqn:MI_s}).
  \textbf{(c)} We estimated the $p$-values of $\mathrm{MI}_{(f_{i})}$ for $f_i \in F$ and $\mathrm{MI}_s$ using permutation testing.
  Each simulation was run 50 times to obtain the MI values capture rates for $\mathrm{MI}_{(f_{i})}$ and $\mathrm{MI}_s$ as the percentage of time where the $p$-values obtained by each method are less than their respective $p$-value thresholds (see Sec.~\ref{subsec:simulation_studies}).
  We then have the capture rate for each of 24 capturing features, as well as the capture rate of $\mathrm{MI}_F$ as a whole -- in a given simulation run, if any of the capturing feature detects the dependence then it is considered captured by $\mathrm{MI}_F$. 
  }
  \label{fig:SimulationDesign}
\end{figure*}

\subsection{Validation}
\label{subsec:simulation_studies}

In the previous section we have introduced the concept of $\mathrm{MI}_f$ as a feature-based MI between two time series, and $\mathrm{MI}_F$ as a method of using a set of candidate features $F$ to compute the $\mathrm{MI}_f$ for each feature $f \in F$ and infer statistical dependence between two time series based on the statistical significance of these feature-based MI values.
In this section, we explore a specific implementation of $\mathrm{MI}_F$ (using a set of 24 candidate time-series features, as detailed later).
We evaluate the behavior and performance of $\mathrm{MI}_s$ and this implementation of $\mathrm{MI}_F$ through a series of experiments on three simulated dynamical processes.
Our aim was to investigate the ability of $\mathrm{MI}_F$ to detect interactions between time series $\mathbf{x}$ and $\mathbf{y}$ for cases in which $\mathbf{y}$ is coupled to a feature of $\mathbf{x}$.

The set-up of these simulation studies is illustrated in Fig.~\ref{fig:SimulationDesign}.
We simulated pairs of source and target time series, where the target time series responds to a feature of the source time series.
We first created a time-series realization $\mathbf{x}$ of the source process $X$.
Next, we generated a feature time series $\mathbf{\tilde{z}}$, by iterating through a sliding window of $\tilde{l}$ samples of $\mathbf{x}$ using a feature function $\tilde{f}$ (note that we use the tilde symbol to denote ground-truth values to distinguish them from the symbols used in the inference of interactions) as
\begin{equation}
\label{eqn:driving_feature}
\tilde{z}_t = \tilde{f}(\mathbf{x}_t^{(\tilde{l})})\,.
\end{equation}
The feature $\mathbf{\tilde{z}}$ was then used to generate time-series data for the target process, hence we refer to $\mathbf{\tilde{z}}$ as the `driving feature', and $\tilde{l}$ as `interaction timescale' throughout the rest of this paper.
The target time series was then generated as a noisy linear transformation of $\mathbf{\tilde{z}}$:
\begin{equation}
\label{eqn:driving_target}
y_{t+1} = (1-\beta) \tilde{z}_{t} + \beta \xi_t\,,
\end{equation}
where $\beta \in [0,1]$ is the noise strength and the i.i.d. noise process follows a normal distribution, $\xi_t \sim N(0,1)$ (note that this parameterization allows us to sample cases with zero noise or zero signal with a finite interval of $\beta$).
Since various features may exhibit distinct ranges of values, we standardized the values in the feature time series $\mathbf{\tilde{z}}$ by $z$-scoring it in all experiments.
This was crucial for fair comparisons among different features, especially when investigating factors like the impact of noise on $\mathrm{MI}_f$.

Having constructed $\mathbf{x}$ and $\mathbf{y}$, with $\mathbf{y}$ responding linearly to a given time-series feature of an interval of the source's past, we proceeded to test how the signal-space $\mathrm{MI}_s$ and feature-space $\mathrm{MI}_F$ methods perform in detecting the interaction between $\mathbf{x}$ and $\mathbf{y}$.
For $\mathrm{MI}_F$, we used a set of 24 candidate capturing features, including the mean, standard deviation, and the 22 features that make up the \textit{catch22} feature set \cite{Lubba2019Catch22}.
We chose \textit{catch22} as a compact set of 22 features that capture different types of time-series properties---including distribution shape, linear and nonlinear autocorrelation structure, extreme event timing among others---allowing us to detect interactions driven by different types of dynamical attributes \cite{Lubba2019Catch22}.
For an overview of these 22 features (including naming details and short descriptions), refer to Table~\ref{tab:FeatureOverview}.
These features were computed over sliding windows of length $l$ samples of $\mathbf{x}$.
When we assumed that we have knowledge of the timescale of interaction, we set $l = \tilde{l}$, otherwise we tested a range of candidate timescales.
Next we computed $\mathrm{MI}_s$ between the source and target time series, and MI between each capturing feature and target time series $\mathrm{MI}_f$.
This was done using the information-theoretic toolkit JIDT \cite{Lizier2014JIDT}.
We chose the Kraskov--St\"ogbauer--Grassberger algorithm \cite{Kraskov2004MI} for estimating MI, as it makes no assumption on the underlying distribution of the input data, can capture nonlinear relationships, and is known to have lower bias and improved data efficiency and accuracy compared to the alternative model-free box-kernel estimation technique \cite{Schreiber2000TE, Lizier2014JIDT}.
In this estimator, the number of nearest neighbours was set to $k = 4$. 
As this estimator is a kernel-based method which has been found to benefit from applying a Theiler window for dynamic correlation exclusion \cite{Cliff2021SignificanceTesting}, we applied a Theiler window here.
In line with common practice \cite{Goswami2019NonlinearTS}, we set the Theiler window of a time series as the first zero-crossing point of the autocorrelation function of the given time series.
The dynamic autocorrelation exclusion window here was set as the maximum of the Theiler windows of the source and target input time series for the estimator (note that in the case of a feature, this was the maximum of the Theiler window of the given feature time series and the Theiler window of the target time series).

We next evaluated the statistical significance of each computed $\mathrm{MI}_f$ and $\mathrm{MI}_s$ value using the method of surrogates.
Surrogate time series are constructed as substitutes for the original time series that preserve key statistical properties, allowing us to generate empirical null distributions for a given hypothesis test \cite{Schreiber2000Surrogates}.
We used surrogate time series to create null distributions for $\mathbf{z}$ (in the case of $\mathrm{MI}_f$) and for $\mathbf{x}$ (in the case of $\mathrm{MI}_s$).
The resulting null distributions were then used to infer the $p$-value of the measures defined as the probability that the surrogate MI measurements are greater than the measured MI.
We applied the Holm--Bonferroni method \cite{Holm1979MultipleTest} to control the family-wise false positive rate for $\mathrm{MI}_F$ when multiple capturing features were used and tested independently.
Note that no such comparison was required for $\mathrm{MI}_s$, as only one hypothesis test is performed.
We used 50 surrogates for each $\mathrm{MI}_s$ measurement and 1000 surrogates for each $\mathrm{MI}_f$ measurement---the greater number of surrogates for $\mathrm{MI}_f$ accounts for the reduced $p$-value resolution due to correcting for multiple testing.
By default, we used random shuffling of the source embedding to generate surrogates when the time-delay embedding dimension, $l$, was small \cite{Lizier2014JIDT}.
For larger embedding dimension $l$, we observed in some of our experiments that this surrogate generation method led to biases in the $p$-value estimation of $\mathrm{MI}_s$.
These biases were not observed when we generated surrogates by circular shifting the source time series (which better preserves autocorrelation) \cite{Vicente2014InformationTransfer, Wibral2014TEinNeuroscience}, when the shifting window is longer than the embedding size.
Hence, for experiments involving longer embedding sizes, $l > 10$, we generated surrogates using the circular shifting method.

In order to evaluate the performance of $\mathrm{MI}_s$ and $\mathrm{MI}_F$ in capturing the dependence between $X$ and $Y$, we ran the simulation 50 times to estimate the dependence `capture rate'.
For $\mathrm{MI}_s$, the capture rate represented the percentage of simulation runs for which $p$-value was less than 0.05.
For each feature $f$ of the 24 capturing features, $F$, the capture rate was calculated as the percentage of simulations for which the $p$-value of $\mathrm{MI}_f$ was less than 0.05 (Holm--Bonferroni adjusted).
The overall capture rate for $\mathrm{MI}_F$ was defined as the percentage of simulations for which at least one feature successfully detected the dependence.
In each simulation run, the samples of the processes will in general differ due to stochasticity in the generative process (including due to random initial conditions and noise).
To ensure reproducibility, we used a fixed random seed for each simulation run.
Given that we simulated source and target processes with feature-mediated interactions, a strong performance should manifest as a higher capture rate here.
To verify the null capture rate, we also used an i.i.d. noise time series $\mathbf{z_\mathrm{null}}$ as a capturing feature.
$\mathbf{z_\mathrm{null}}$ is simulated by generating samples from a normal distribution, $\mathcal{N}(0,1)$, as:
\begin{equation}
\label{eqn:znull}
(z_\mathrm{null})_t \sim \mathcal{N}(0,1)\,.
\end{equation}

\begin{figure*}
    \centering
  \includegraphics[width=0.9\textwidth]{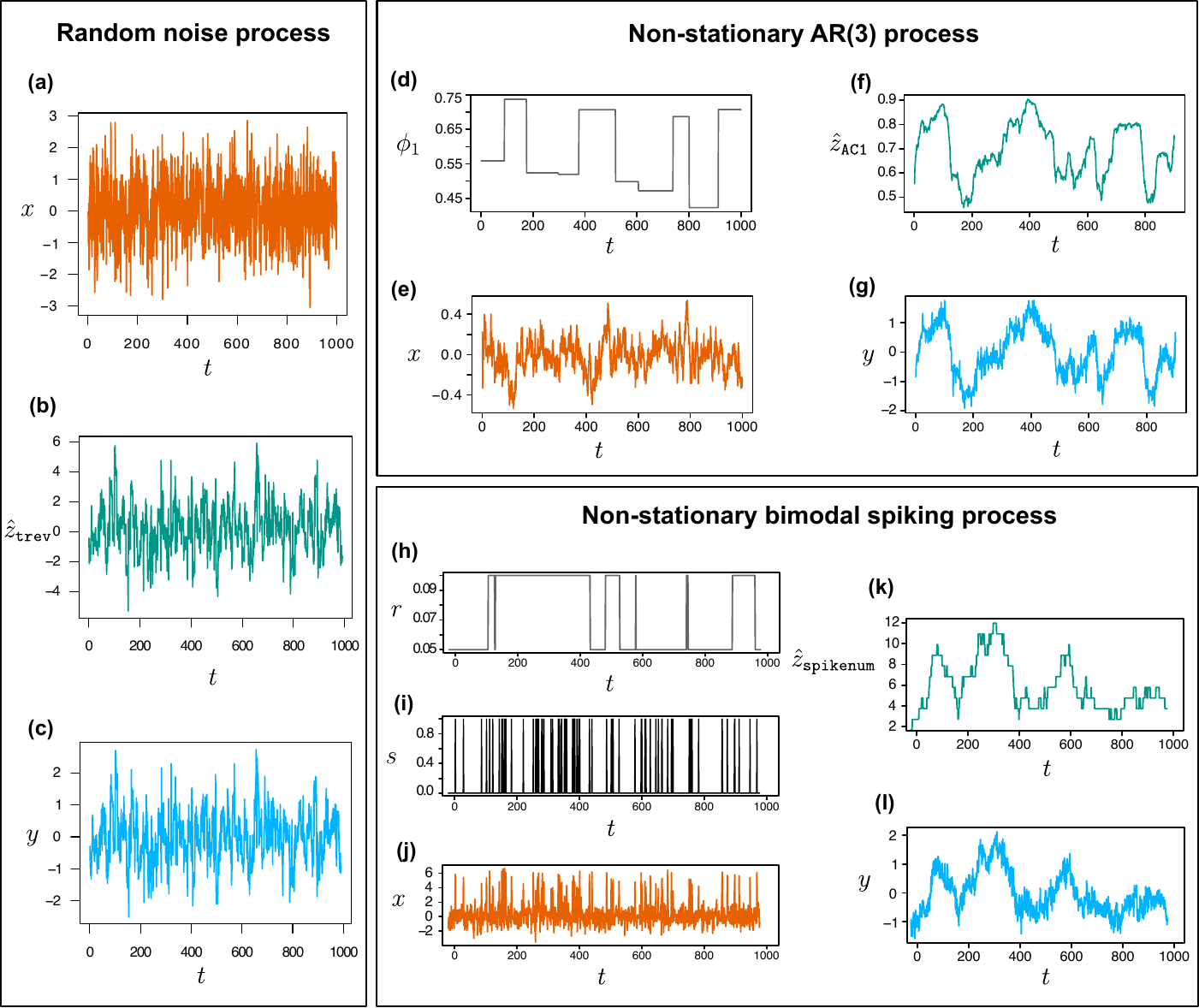}
  \caption{
  \textbf{We studied three generative source processes, with target processes linearly coupled to local dynamical properties of each source process.}
  For each process, here we show example time-series realizations. 
  The plots represent time series of length 1000 for each process.
  \textbf{(a)--(c)}
  A random noise process, \cref{eqn:random_noise}.
  \textbf{(a)} A time-series realization of the the process.
  \textbf{(b)} An example feature time series, for the feature labeled \texttt{trev}, $\mathbf{\tilde{z}_{\texttt{trev}}}$ (defined in Eq.~\eqref{eqn:trev}), which computes the average across the time series of the cube of successive time-series differences, computed over sliding windows of duration $\tilde{l} = 10$ of $\mathbf{x}$ (cf. Eq.~\eqref{eqn:driving_feature}).
  \textbf{(c)} The simulated target process, $\mathbf{y}$ (a noisy linear function of $\mathbf{\tilde{z}_{\texttt{trev}}}$, as defined in Eq.~\eqref{eqn:driving_target}).
  \textbf{(d)--(g)} 
  A non-stationary third-order auto-regressive, AR(3), source process, with a time-varying lag-1 coefficient, $\phi_1$, Eq.~\eqref{eqn:ar3}.
  We plot example realizations of:
  \textbf{(d)} the  $\boldsymbol{\phi_{1}}$ time series, generated by a piece-wise constant function [see \cref{eqn:ar3_phi1}],
  \textbf{(e)} the source time series, $\mathbf{x}$,
  \textbf{(f)} the driving feature that mediates the coupling to the target process, lag-1 autocorrelation $\mathbf{\tilde{z}_{{\texttt{AC1}}}}$ (computed over sliding windows of size duration $\tilde{l} = 100$ of $\mathbf{x}$),
  \textbf{(g)} the target time series, $\mathbf{y}$ [cf. Eq.~\eqref{eqn:driving_target}].
  \textbf{(h)--(l)}
  A non-stationary bimodal spiking process, characterized by its spike rate, switches between two values over time [see \cref{eqn:bimodal_spike_rate_switch,eqn:bimodal_spike_rate,eqn:bimodal_spike_x} for how this process is defined mathematically and how the time-series samples are generated].
  We plot example realizations of:
  \textbf{(h)} The spike rate time series $\mathbf{r}$ takes one of two values, $0.05$ or $0.1$, where the switching between the two spike rate values is governed by a Bernoulli distribution with probability $p$ [see Eqs.~\eqref{eqn:bimodal_spike_rate_switch} and ~\eqref{eqn:bimodal_spike_rate}].
  \textbf{(i)} The spike time series, $\mathbf{s}$, indicates the presence (1) or absence (0) of a spike at each time point, and is generated by sampling from a Bernoulli distribution with probability $\mathbf{r}$.
  \textbf{(j)} The source time series $\mathbf{x}$ is generated from $\mathbf{s}$ through a noisy rescaling of the spike time series $\mathbf{s}$, as given by \cref{eqn:bimodal_spike_x}.
  \textbf{(k)} We chose the driving feature as the number of spikes, $\mathbf{\tilde{z}_{\texttt{spikenum}}}$ (defined in as the number of times over a time-series window of length $\tilde{l}$ that the sample value exceeds a threshold of four),
  \textbf{(l)} The target process, $\mathbf{y}$, is a noisy linear function of $\mathbf{\tilde{z}_{\texttt{spikenum}}}$, as Eq.~\eqref{eqn:driving_target}.
  } 
  \label{fig:GenerativeProcesses}
\end{figure*}

\subsection{Simulated processes for validation}
\label{subsec:generative_processes}
In the previous section we outlined our methodology for testing $\mathrm{MI}_F$ through simulation studies. 
Here we discuss in detail the dynamical processes used in our simulation studies.
A visual representation is illustrated in Fig.~\ref{fig:GenerativeProcesses}, which shows the simulated source processes, their driving features, and corresponding target processes.

We selected generative processes and interaction timescales $\tilde{l}$ with the aim of examining the ability of $\mathrm{MI}_F$ to infer long-timescale interactions, which we hypothesized would be more challenging for $\mathrm{MI}_s$ to detect.
We first studied the simple case of a stationary noise process to test the ability of $\mathrm{MI}_F$ and $\mathrm{MI}_s$ to detect coupling driven by dynamical properties that do not change over long periods of time.
Then we explored non-stationary processes governed by parameters that change over time in which the target process responds linearly to a time-series feature that closely resembles the time-varying parameter of the non-stationary source process.
This approach resembles real-world settings where the target process responds to the inferred statistical properties from the source process time series, rather than having direct knowledge of the source parameters.
In such interactions, entities respond to discernible patterns and statistical regularities without direct insight into internal workings.
For instance, a financial market adapts to observed patterns in economic indicators without explicit knowledge of underlying economic parameters.

\paragraph{Random noise.}
We first aimed to investigate the performance of $\mathrm{MI}_F$ relative to $\mathrm{MI}_s$ in the case where the target responds only to statistical fluctuations in a noise process.
To achieve this, we simulated a random noise process by generating i.i.d. samples from a normal distribution, $\mathcal{N}(0,1)$, as
\begin{equation}
\label{eqn:random_noise}
x_t \sim \mathcal{N}(0,1)\,.
\end{equation}
An example realization, $\mathbf{x}$, is plotted in Fig.~\ref{fig:GenerativeProcesses}(a).
To study how well $\mathrm{MI}_F$ and $\mathrm{MI}_s$ could capture a source--target interaction mediated by a given time-series feature, we tested the case in which the target responds to each feature in our candidate feature set.
An example is shown for the case where the driving feature is the time-reversibility statistic $\texttt{trev}$ (from \textit{catch22} \cite{Lubba2019Catch22}, see Table~\ref{tab:FeatureOverview}) and for a sliding window duration $\tilde{l} = 10$ in Fig.~\ref{fig:GenerativeProcesses}(b), for the resulting driving feature time series, $\mathbf{\tilde{z}}_{\texttt{trev}}$.
Concretely, the driving feature $\mathbf{\tilde{z}_{\texttt{trev}}}$ (see Fig.~\ref{fig:GenerativeProcesses}(b)) was computed as:
\begin{equation}
\label{eqn:trev}
 ({\tilde{z}_{\texttt{trev}})_t}  = \frac{1}{10} \sum_{i=t-9}^t (x_{i+1}-x_i)^3\,.
\end{equation}
We then standardized $\mathbf{\tilde{z}_{\texttt{trev}}}$ and generated the target $\mathbf{y}$ from Eq.~\eqref{eqn:driving_target}, as shown in Fig.~\ref{fig:GenerativeProcesses}(c).

\paragraph{A non-stationary third-order auto-regressive process, AR(3).}\label{ar3Process}
We simulated a feature-based source--target interaction in which the source dynamics are governed by a non-stationary AR$(3)$ process.
This type of non-stationary auto-regressive process, in which the auto-regressive coefficients change over time, are common in many fields, as they can capture evolving dependencies \cite{Leybourne1996NonstationarityTSEconomics, Cassidy2002BayesianARModelBiomed}.
Here we chose to study an AR(3) process \cite{BoxJenkins76} with a time-varying lag-1 autoregressive coefficient as a specific case of AR($p$) processes \cite{BoxJenkins76} to evaluate our method in dynamic scenarios commonly encountered in practice.
A time-series realization, $\mathbf{x}$, of this source process is given by
\begin{equation}
\label{eqn:ar3}
 x_t = (\phi_1)_t x_{t-1} + \phi_2 x_{t-2} + \phi_3 x_{t-3} + \gamma \xi_t\,, 
\end{equation}
where we set the lag-2 coefficient $\phi_2 = 0.1$, lag-3 coefficient $\phi_3 = 0.1$, noise strength $\gamma = 0.1$, and $\xi_t \sim N(0,1)$ are i.i.d. samples from a Gaussian distribution.
The values for $\phi_2$ and $\phi_3$ were chosen to be smaller than $\phi_1$ to prevent interactions mediated by changes in $\phi_1$ from being overshadowed by the impact of $\phi_2$ and $\phi_3$ on the sample values.
To generate a non-stationary process, we used a piece-wise constant function to generate time-series values for the lag-1 coefficient, $\boldsymbol{\phi}_{1}$, as:
\begin{equation}
\label{eqn:ar3_phi1}
(\phi_1)_t = \begin{cases} 
  \varphi_1 & \text{if } t \leq t_{1}\,, \\
  \varphi_2  & \text{if } t_{1} < t \leq t_{2}\,, \\
  \vdots & \\
  \varphi_k  & \text{if } t_{k-1} < t \leq t_{k}\,, \\
  \vdots & \\
  \varphi_n & \text{if } t_9 < t \leq T\,, \\
\end{cases}
\end{equation}
where each $\varphi_k \sim \mathcal{U}(0.3, 0.8)$ was randomly sampled from a uniform distribution in the range 0.3 to 0.8, and $t_1$, $t_2$, ..., $t_9$ partition the time sequence into 10 irregular intervals.
The durations of these intervals were generated using a simple heuristic: randomly selecting 10 values from a normal distribution $N(\mu,\sigma)$ where $\mu = T/10$ and $\sigma = 20$, and then rounding the results to the nearest positive integers.
Adjustments, by increments of $\pm 1$ as necessary, were made until reaching a cumulative sum equal to $T$.
The range of values, $0.3 \leq (\phi_1)_t \leq 0.8$ was chosen to be within the limits that ensure the stability of $\mathbf{x}$.
The choice of 10 values for $\boldsymbol{\phi}_{1}$ ensured some variability and sufficient dwell time at each value.
Figure~\ref{fig:GenerativeProcesses}(d) plots an example realization of $\boldsymbol{\phi}_{1}$ alongside a corresponding time-series realization $\mathbf{x}$ in Fig.~\ref{fig:GenerativeProcesses}(e).

We then simulated a scenario that aimed to mimic a target process responding to non-stationary variation in the statistics of a source time series (governed by a time-varying parameter).\
Here we chose the driving feature (of the source process to which the target process responds) as the lag-1 autocorrelation \cite{Veneables2002ModernStats}, with the target time series, $\mathbf{y}$, generated as a noisy linear response as usual [as Eq.~\eqref{eqn:driving_target}].
A realization of the lag-1 autocorrelation feature, $\mathbf{\tilde{z}_\mathrm{AC1}}$, of $\mathbf{x}$, and its corresponding target time series $\mathbf{y}$, are illustrated in Figs~\ref{fig:GenerativeProcesses}(f) and (g), respectively.
We can see from Figs~\ref{fig:GenerativeProcesses}(d) and (f) that $\mathbf{\tilde{z}_\mathrm{AC1}}$ loosely aligns with the variations in $\boldsymbol{\phi_{1}}$ values, while $\mathbf{y}$ is just a noisier, scaled version of $\mathbf{\tilde{z}_\mathrm{AC1}}$.
        
\paragraph{Bimodal spiking process.}
We next analyzed a non-stationary source process characterized by a time-varying spike rate.
This choice of source process finds inspiration in real-world scenarios, particularly in neuroscience, where interactions can often be driven by the precise timing of spikes or neuronal firing rates \cite{Feldman2012STDP, Usrey2000SynapticInteractions, Gerstner1997NeuralCodes}.
Our simulation emulates a communication strategy in which one neuron influences another by varying its spike rate, and the target neuron requires an extended time windows to decipher the varying spike rate of the source neuron.
This scenario mirrors real-world situations in which a target process (a system with memory and computational capabilities) captures and responds to underlying variations of an observed source process on longer timescales.

We generated a process with alternating spike rates: the spike-rate time series, $\mathbf{r}$, over a time window of 100 samples, switches between two values, $\lambda_1 = 0.05$ and $\lambda_2 = 0.1$.
An example time series for $\mathbf{r}$ is plotted in Fig.~\ref{fig:GenerativeProcesses}(h).
A parameter, $p = 0.01$, determines the probability of switching between these two values of spike rate at each time step, yielding a binary indicator time series, $\mathbf{c}$: $c_t = 1$ denotes a switch of rate (from $\lambda_1$ to $\lambda_2$, or vice-versa), while $c_t = 0$ retains the current rate.
$\mathbf{c}$ follows a Bernoulli distribution with probability $p$:
\begin{equation}
\label{eqn:bimodal_spike_rate_switch}
c_t \sim B(1,p) \,.
\end{equation}
The spike rate dynamics follows
\begin{equation}
\label{eqn:bimodal_spike_rate}
r_t = (1 - c_t)r_{t-1} + c_t(\lambda_1 + \lambda_2 - r_{t-1})\,.
\end{equation}
We then generated the spike time series, $\mathbf{s}$, where at each time point, $s_t = 1$ (representing a spike) with probability $r_t$ and $s_t = 0$ with probability $1-r_t$ (representing an absence of a spike).
The source time series, $\mathbf{x}$, was then constructed as a noisy rescaling of the spike time series,
\begin{equation}
\label{eqn:bimodal_spike_x}
x_t = \delta s_t + \xi_t\,,
\end{equation}
where $\delta = 5$, and $\xi_t \sim N(0,1)$ are i.i.d. samples from a Gaussian distribution.
The relatively high value of $\delta$ compared to the mean value of $\xi_t$ is intended to create the jump in values (or `spikes') in $\mathbf{x}$.
The spike indicator time series $\mathbf{s}$ and the source time series $\mathbf{x}$ are plotted in Figs~\ref{fig:GenerativeProcesses}(i) and (j), respectively.
We can see that the spikes in $\mathbf{x}$ correspond to times at which $s_t = 1$.

As this process is characterized by a varying rate of spikes, we chose the number of spikes over a time-series window of $\tilde{l}$ samples as the driving feature in experiments using this process.
A spike here was defined as a sample $x_t$ where $x_t > 4$; we chose four as the threshold for spike detection due to the parameter values used in \cref{eqn:bimodal_spike_x}.
The driving feature $\mathbf{\tilde{z}}_\texttt{spikenum}$ is illustrated Fig.~\ref{fig:GenerativeProcesses}(k) where we can observe that the high values in $\mathbf{\tilde{z}_\texttt{spikenum}}$ are aligned with the periods of higher spike densities in $\mathbf{x}$ [shown in Fig.~\ref{fig:GenerativeProcesses}(j)].
The target time series $\mathbf{y}$ generated from $\mathbf{\tilde{z}}_\texttt{spikenum}$ is plotted in Fig.~\ref{fig:GenerativeProcesses}(l); we can see that it is a noisy scaled version of $\mathbf{\tilde{z}_\texttt{spikenum}}$.

\section{Results}
\label{sec:results}

This study aims to investigate the ability of $\mathrm{MI}_F$ to detect underlying feature-mediated dependencies between two processes across various levels of noise, interaction timescales, and time-series lengths.
In each case, we compare the ability of $\mathrm{MI}_F$ to detect simulated feature-mediated source--target dependencies to that of $\mathrm{MI}_s$.
First, in Sec.~\ref{subsec:ValidationSimpleSettings}, we verify the behavior of $\mathrm{MI}_F$ on a simple case in which the source process is a random noise process.
Then, in Sec.~\ref{subsec:Robustness}, we study how the method performs when there are non-stationary dynamics in the source, and evaluate the robustness of the methods across variation in noise levels and interaction timescales.

\begin{figure*}[ht!]
    \centering
  \includegraphics[width=\textwidth]{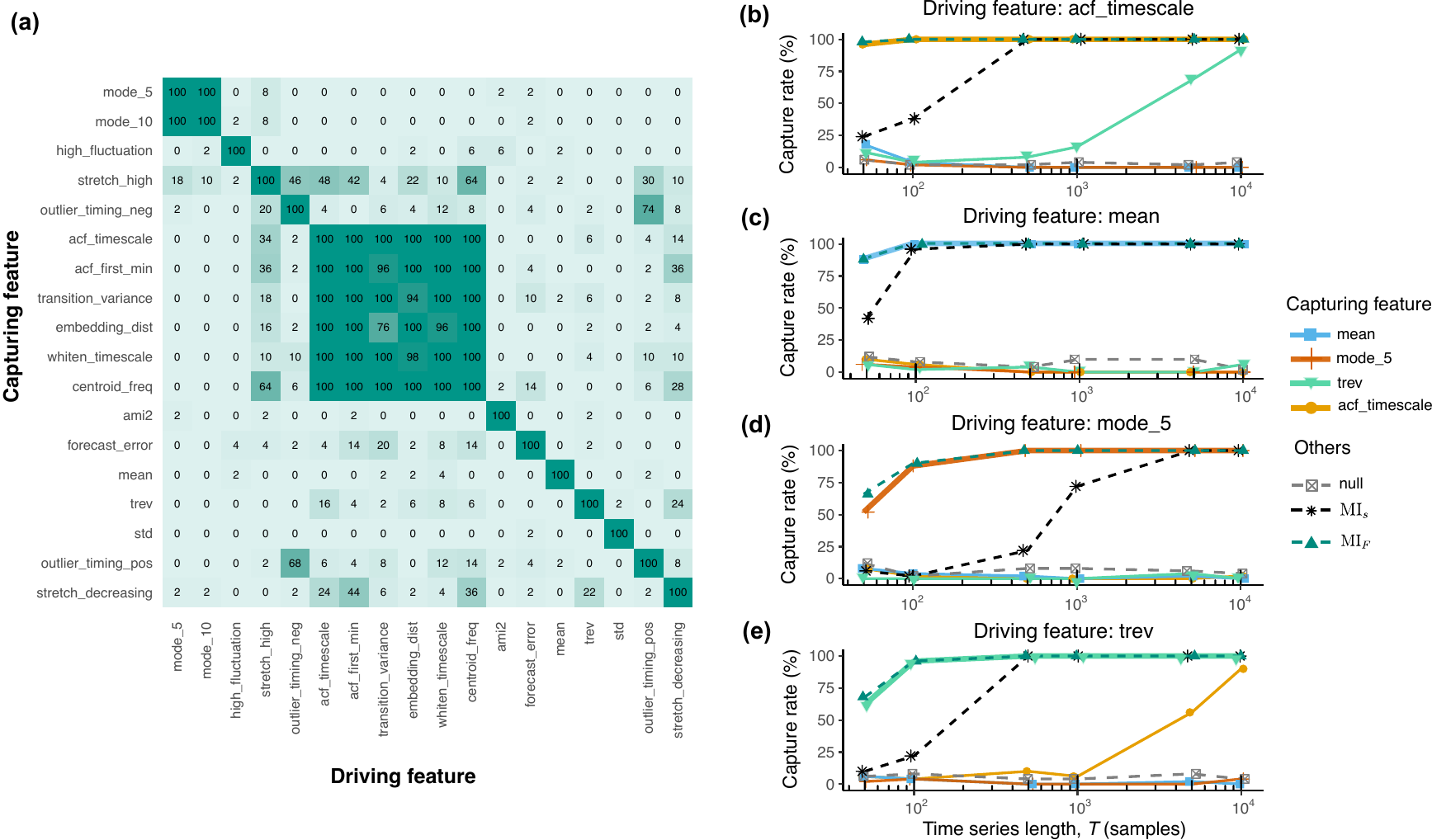}
  \caption{
  \textbf{$\mathrm{MI}_F$ can detect feature-mediated interactions from time-series data more efficiently than $\mathrm{MI}_s$ when the capturing feature set includes the driving feature, shown here for simulations using a noisy source process.}
  \textbf{(a)}
  We simulated interactions mediated by each of 24 driving time-series features \cite{Lubba2019Catch22} and used $\mathrm{MI}_F$ with 24 capturing features on each simulation.\
  Features with zero standard deviation (or very few distinct values) were removed from the analysis, resulting in the 18 features shown here.
  This matrix displays the capture rate for each combination of capturing feature (rows) and driving feature (columns) for a random noise process [Eq.~\eqref{eqn:random_noise}] for time series of length $T = 1000$, interaction timescale $\tilde{l} = 10$ samples, and noise strength $\beta = 0.2$.
  The time-series features are ordered by their similarity to each other, measured by the correlation of feature values across time series for each pair of features.
  Cells are colored by the capture rate of MI$_f$ for each combination of capturing and driving feature (darker colors indicate higher capture rates).
  \textbf{(b)--(e)}
  The capture rates for four selected features---autocorrelation timescale (\texttt{acf\_timescale}), a time-reversibility metric (\texttt{trev}), and histogram mode (\texttt{mode\_5}), and mean (\texttt{mean})---are plotted as a function of time-series length $T$ (note the logarithmic scale for $T$).
  Each plot shows the variation in capture rates for a given driving feature:
  \textbf{(b)} \texttt{acf\_timescale},
  \textbf{(c)} \texttt{mean},
  \textbf{(d)} \texttt{mode\_5}, and
  \textbf{(e)} \texttt{trev}.
  Dashed dark green lines represent the capture rate of $\mathrm{MI}_F$ using the full set of 24 candidate features (see Sec.~\ref{subsec:simulation_studies}), i.e., if any features among the 24 capturing features detects a statistical dependence (controlling the family-wise error at $\alpha = 0.05$) then it is considered detected by $\mathrm{MI}_F$.
  Dashed black lines represent the capture rate of $\mathrm{MI}_s$. 
  Dashed grey lines represent the capture rate from using noise time series $\mathbf{z_\mathrm{null}}$ [defined in Eq.~\eqref{eqn:znull}].
  Thick lines distinguish cases for which the capturing feature is the same as the driving feature.
  }
  \label{fig:RandomProcessResults}
\end{figure*}

\subsection{Performance on a stationary noise source process}
\label{subsec:ValidationSimpleSettings}

We first studied the performance of $\mathrm{MI}_F$ in an idealized setting: a stationary noise source process, a short interaction timescale ($\tilde{l} = 10$), and a low noise level ($\beta = 0.2$), with the objective of examining the performance of $\mathrm{MI}_F$ in this test case.
Note that, in this case, variations in the statistics of the source process from window to window are due to random sampling (non-stationary variation is studied later).
We aimed to address three primary questions:
(1) In the ideal case where $\mathrm{MI}_F$ has access to the driving feature and interaction timescale (i.e., $z_j = \tilde{z}$ and $\tilde{l} = l = 10$ ), can it successfully detect the feature-mediated dependence?;
(2) Can alternative features be used to detect dependence mediated by $\tilde{z}$?; and
(3) How effective is $\mathrm{MI}_F$ at capturing this dependence?

We used each feature from our set of 24 time-series features as driving features in turn, thus testing 24 potential types of source--target feature-based interactions.
We then tested the ability of each of those 24 features (treated as capturing features) to detect the pairwise interaction in each case (correcting for 24 independent comparisons using the method of Holm--Bonferroni \cite{Holm1979MultipleTest}, see \textit{Methods}).
Using each feature in our candidate feature set as both a driving and capturing feature served a dual purpose for evaluating MI$_f$:
(i) testing its capture rate when the capturing feature matches the driving feature; and
(ii) testing its ability to detect a source--target dependence when the capturing feature differs from the driving feature.
The latter is particularly important in real-world scenarios where we do not typically have prior knowledge of the precise driving feature, and it may be outside of capturing set $F$.

Results are shown as a matrix of capture rates for each capturing feature (row) against each driving feature (column) in Fig.~\ref{fig:RandomProcessResults}(a) (features with zero standard deviation or with very few distinct values are removed from the analysis, leaving 18 features shown here).
Note that features have been reordered to place similar features close to each other (using complete linkage hierarchical clustering based on Spearman correlations).
First we verified that, in the ideal case when a capturing feature matches the driving feature, its $\mathrm{MI}_f$ can capture the dependence well, which should result in high capture rates along the diagonal line of the matrix.
Indeed, all capture rates along the diagonal are at 100\%, indicating strong performance of $\mathrm{MI}_f$ in capturing dependence when the capturing feature aligns with the driving feature.
In addition to this diagonal structure, we also observed noticeable groups of features with high rates of capturing a source--target interaction mediated by each other.
This suggests that a feature highly correlated with another (as suggested by their proximity in the matrix) can indeed capture dependence mediated by the other feature.
The most prominent example is a large cluster of inter-related features that capture different aspects of signal self-correlation properties, including statistical properties of the Fourier spectrum (like centroid frequency \verb|centroid_freq|) and properties of the autocorrelation function (such as the first $1/e$ crossing of the ACF \verb|acf_timescale| and the first minimum of the ACF \verb|acf_first_min|).
Another example is two histogram-shape features (position of the mode using 10 bins, \verb|mode_10|, versus 5 bins, \verb|mode_5|) which can be used to capture dependence caused by each other.
We also note that, while the matrix is mostly symmetric, there are some cases of substantial asymmetry.
As the matrix shows the percentage of simulation runs where the MI between the capturing feature and a noisy linear scaling of the driving feature is statistically significant, these asymmetries can mostly be accounted for by both the asymmetry of introducing noise in one variable versus the other, as well as the stochasticity of $p$-value estimation from a finite surrogate ensemble.
Overall, these results demonstrate that $\mathrm{MI}_F$ performs well when the capturing feature matches the driving feature and that correlated alternative capturing features can also effectively capture dependence mediated by the driving feature.

Next, we assessed the effectiveness of $\mathrm{MI}_f$ for each selected feature as well as $\mathrm{MI}_F$ as a whole in capturing source--target dependence by examining its capture rate as a function of the time-series length, $T$.
We conducted tests across a range of time-series lengths, $T = 50, 100, 500, 1000, 5000$ and $10,000$ samples.
We also computed the capture rate of $\mathrm{MI}_s$ for comparison.
The capture rates are plotted as a function of $T$ on a logarithmic scale for four exemplary driving features in Figs~\ref{fig:RandomProcessResults}(b)--(e).
These four selected driving features were chosen from different classes of features to illustrate the performance of $\mathrm{MI}_s$ and $\mathrm{MI}_f$ at detecting a dependence mediated by various types of statistical properties, from distributional properties (the mean, Fig.~\ref{fig:RandomProcessResults}(c); and 5-bin histogram mode, \verb|mode_5|, Fig.~\ref{fig:RandomProcessResults}(d)) to autocorrelation-based properties (the first $1/e$ crossing of the ACF, \verb|acf_timescale|, Fig.~\ref{fig:RandomProcessResults}(b); and time reversibility measure, \verb|trev|, Fig.~\ref{fig:RandomProcessResults}(e)).

We first note that the capture rate for the noise time series $\mathbf{z_\mathrm{null}}$ was approximately 5\% for all experiments and across all time-series lengths, $T$, aligning with the false positive rate.
When using the driving feature as the capturing feature, the resulting $\mathrm{MI}_f$ exhibits a very high capture rate for time-series lengths $T \gtrapprox 100$ samples, which increases with $T$.
Interactions mediated by some features, such as \verb|acf_timescale| and \verb|mean|, were detected by $\mathrm{MI}_F$ from very short time series ($T = 50$) with capture rates over 85\%.
Among the features selected here, when the driving feature is \verb|acf_timescale| [Fig.~\ref{fig:RandomProcessResults}(b)] or \verb|trev| [Fig.~\ref{fig:RandomProcessResults}(e)], other capturing features could also detect the dependence caused by a different driving feature when given enough samples ($T = 10,000$ samples in this case), consistent with the results in Fig.~\ref{fig:RandomProcessResults}(a).

By contrast, $\mathrm{MI}_s$ exhibited a considerably lower capture rate than $\mathrm{MI}_F$ when $\mathrm{MI}_F$ has the driving feature in the candidate feature set $F$ for shorter time-series lengths up to $T = 1000$ samples.
For instance, for time series of length $T = 50$ samples, the capture rate of $\mathrm{MI}_s$ for \verb|mode_5| and \verb|trev| are under 10\%.
Even at $T = 100$ samples, the capture rate is still under 50\% for all the features except \verb|mean|.
While the performance of $\mathrm{MI}_s$ improves with $T$, the improvement varied among driving features, improving rapidly when the interaction is mediated by a single feature like the mean [a capture rate over 90\% with $T = 100$ samples, Fig.~\ref{fig:RandomProcessResults}(c)], but slowly when the interaction is mediated by a more complex feature like the histogram mode \verb|mode_5|, where $T = 5000$ is required to teach a similarly high capture rate.
This difference in the detection power of $\mathrm{MI}_s$ for these two driving features may be due to their differing complexity: \verb|mean| is a smooth rolling average of the time series (a linear function on $\mathbf{X}_t^{(l)}$), making it relatively easy for the MI estimator to infer dependence.
By contrast, \verb|mode_5| is a more complex feature, making the dependence more challenging to infer from a small number of samples, $T$.

In summary, our experiments using a random source process demonstrate that $\mathrm{MI}_F$ can detect a feature-mediated dependency in idealized settings of low noise, having knowledge of the interaction timescale $\tilde{l}$, and when our capturing feature set includes the driving feature $\tilde{z}$.
We also showed that features with similar behavior can be used to infer dependence caused by another.
In this advantageous setting, where we have the driving feature amongst candidate features, $\mathrm{MI}_F$ is more sensitive than $\mathrm{MI}_s$, being able to detect an underlying feature-mediated dependence from much shorter time series.

\subsection{Performance in settings with non-stationary source processes}
\label{subsec:Robustness}

In the previous section, we investigated the simple case where the target process responds to the window-to-window statistical fluctuations of a stationary noisy source process.
However, real-world systems often involve more intricate dynamics, where the target responds to meaningful statistical changes in a non-stationary source process over time.
Note that here we adopt a pragmatic definition of non-stationarity as variation over long periods of time of statistical properties of a time series \cite{Schreiber1997DetectingNonstationarity, Aguirre2012Nonstationarity}.
We study two such systems in this section, simulated by adding time variation to a key parameter governing the source dynamics (see Sec.~\ref{subsec:generative_processes}): the first is an AR(3) process, where the lag-1 coefficient $\phi_1$ varies over time, and the second is a bimodal spiking process where the parameter governing the spike rate changes over time.
In each case, the target process responds to a statistical property of the source process that is induced by the time-varying parameter: in the first case, the target responds to the lag-1 autocorrelation feature $\mathbf{\tilde{z}_\mathrm{AC1}}$, which is sensitive to changes in $\phi_1$; and in the second case the target responds to the number of times $x_t > 4$ in a recent time window ($\mathbf{\tilde{z}}_\texttt{spikenum}$). 
This matches the situation where the target is meaningfully responding to some underlying parameter governing changes in the statistics of the source process, but without direct access to that underlying parameter (only through observations of the source time series).
We aimed to investigate whether $\mathrm{MI}_F$ can detect this underlying source--target interaction in the case that $\mathrm{MI}_F$ has access to a set of features including those sensitive to the driving feature.
We further study the performance of $\mathrm{MI}_F$ as a function of the noise strength $\beta$, interaction timescale $\tilde{l}$, and capturing timescale $l$.

\subsubsection{Robustness to noise}
\label{sec:noise_robustness}
\begin{figure*}[ht!]
    \centering
  \includegraphics[width=\textwidth]{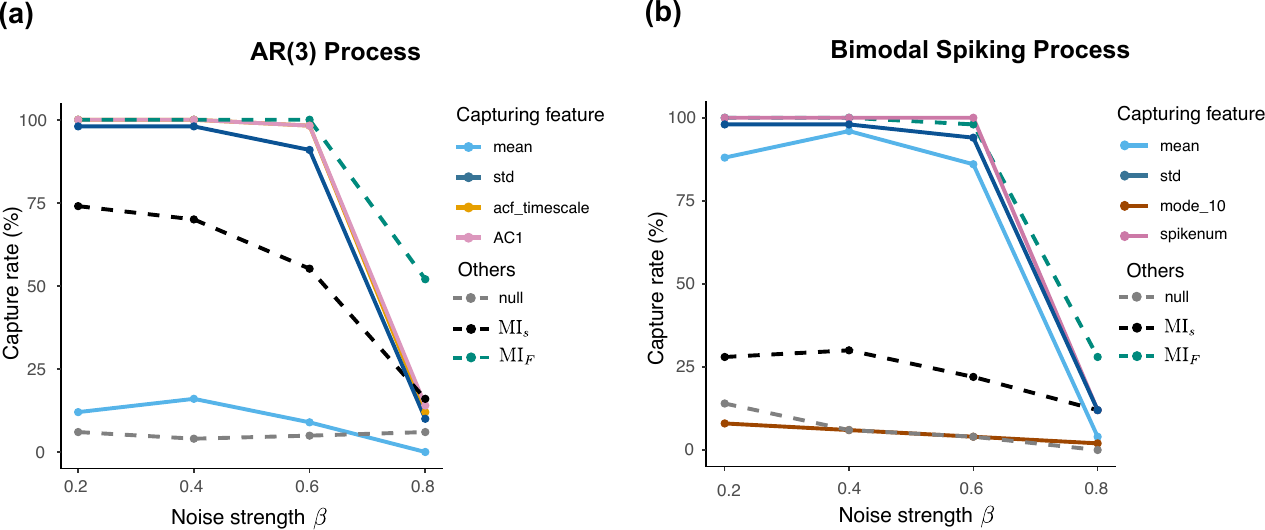}
  \caption{
  \textbf{$\mathrm{MI}_F$ is more robust to noise than $\mathrm{MI}_s$.}
  We evaluated the performance of $\mathrm{MI}_F$ and $\mathrm{MI}_s$ on various levels of noise strength for a time series of length $T = 1000$ samples and interaction timescale $\tilde{l} = 100$ samples, for two processes:
  \textbf{(a)} an AR(3) process, \cref{eqn:ar3}, where the driving feature is the lag-1 autocorrelation; and
   \textbf{(b)} a bimodal spiking process, \cref{eqn:bimodal_spike_x}, where the driving feature is the number of spikes [defined in Sec.~\ref{subsec:generative_processes}(c)]. 
  Capture rates are plotted as a function of the noise strength, across a range $0.2 \leq \beta \leq 0.8$ for four selected capturing features.
  The pink lines illustrate the capture rate when the capturing feature is the driving feature.
  Dashed dark green lines represent the capture rate of $\mathrm{MI}_F$ using the set of 24 candidate features (see Sec.~\ref{subsec:simulation_studies}).
  Dashed black lines represent the capture rate of $\mathrm{MI}_s$. 
  Dashed gray lines represent the capture rate from using noise time series $\mathbf{z_\mathrm{null}}$.
  }
  \label{fig:NoiseAnalysis}
\end{figure*}

In this section, we explore the impact of measurement noise of the target response on the performance of $\mathrm{MI}_F$ and $\mathrm{MI}_s$, by varying the noise strength $\beta$ [Eq.~\eqref{eqn:driving_target}].
$\mathrm{MI}_s$, which operates in the signal space of time-series data, faces a significant limitation in its vulnerability to additive measurement noise, particularly when inferring relationships in high-dimensional spaces due to long-timescale interactions.
Assuming that we have a relevant capturing feature set, we hypothesized that by evaluating each of the capturing features as mediating the source--target interaction, $\mathrm{MI}_F$ should effectively capture and integrate information within that time interval, resulting in a more efficient representation of the process dynamics that is less sensitive to noise.
We studied the impact of noise strength $\beta$ for a time-series length $T = 1000$ samples for both the AR(3) and the bimodal spiking processes in this section.
Since the dynamics of both processes are characterized by an underlying parameter varying at an average rate of 0.01 per sample, we matched the rate of parameter change by setting the  interaction timescale $\tilde{l} = 100$.
This choice aims to simulate scenarios where the target process possesses accurate knowledge of the timescale governing changes in the underlying parameter of the source process (note that we will relax this assumption in the subsequent section).
Here we also assume that we have knowledge of the interaction timescale, and set $l = \tilde{l} = 100$.
This assumption will also be relaxed in a later section, where we investigate how $\mathrm{MI}_F$ and $\mathrm{MI}_s$ perform without precise knowledge of the interaction timescale.

The results are presented in Fig.~\ref{fig:NoiseAnalysis}, showing the capture rates of $\mathrm{MI}_F$ along with those for the $\mathrm{MI}_f$ of the selected capturing features. 
We also show $\mathrm{MI}_s$, and use a noise time series $\mathbf{z}_\mathrm{null}$ to verify the false positive capture rate [Eq.~\ref{eqn:znull}].
The features selected for illustration here include mean (\verb|mean|), standard deviation (\verb|std|), and features conceptually related to the driving feature---such as autocorrelation timescale (\verb|acf_timescale|), an autocorrelation-related property sensitive to lag-1 autocorrelation (\verb|AC1|), and 10-bin histogram mode (\verb|mode10|), a distributional property that may be sensitive to the number of spikes (\verb|spikenum|).
We also plot the capture rate given by the driving feature---\verb|AC1| for the AR(3) process, and \verb|spikenum| for the bimodal spiking process---for comparison.
The full set of results with the capture rate of each feature in the feature set, along with the capture rate of $\mathrm{MI}_s$ and $\mathbf{z}_\mathrm{null}$ are included in the supplementary tables 1--6.
First, we verified that the capture rate of $\mathbf{z}_\mathrm{null}$ for both processes is around 5\%, matching the false-positive capture rate.
We found that $\mathrm{MI}_F$ exhibits strong performance for $\beta \leq 0.6$ in the range explored for both processes, with capture rate of 100\% for the AR(3) process and $ \geq 98\%$ for the bimodal spiking process, respectively.
By contrast, $\mathrm{MI}_s$ did not perform well for either process when $\beta \geq 0.4$ and its performance decreased with increasing noise levels.
For the AR(3) process, this degradation with increasing noise was quite pronounced, decreasing from 70\% at a noise strength of $\beta = 0.4$ to 52\% at $\beta = 0.6$, and further to 16\% at $\beta = 0.8$.

Promisingly, some features could detect the interaction mediated by a different feature, consistent with the findings from the experiments with random noise process (Fig.~\ref{fig:RandomProcessResults}).
For example, for the AR(3) process [Fig.~\ref{fig:NoiseAnalysis}(a)], the autocorrelation timescale (labeled \verb|acf_timescale|) and standard deviation (labeled \verb|std|) detected the dependence mediated by the driving feature, lag-1 autocorrelation, with capture rates of 98\% and 91\% respectively, even at noise levels up to $\beta = 0.6$.
The reason is clear in the case of autocorrelation timescale (\verb|acf_timescale|), which captures a similar dynamical property to lag-1 autocorrelation, making it an effective proxy for inferring the interaction.
In the case of standard deviation, the connection is less obvious.
We discovered from simulations that an increase in the underlying parameter $\phi_1$ governing $\bold{x}$ leads to an increase in the variance of $\bold{x}$.
Given that the driving feature \verb|AC1| closely tracks $\phi_1$, the variance of  $\bold{x}$ then becomes sensitive to \verb|AC1|.
For the bimodal spiking process [Fig.~\ref{fig:NoiseAnalysis}(b)], a greater number of spikes in a given window affects the distribution of values (including increasing the mean and standard deviation of the windowed time series) such that the candidate features, mean (\verb|mean|) and standard deviation (\verb|std|), could effectively detect an interaction mediated by the spike count (\verb|spikenum|), with capture rates of 86\% and 94\% respectively, even up to high noise levels $\beta = 0.6$.
These features could effectively capture the dependence, as the presence of spikes significantly alter the mean and standard deviation of an observed window, making these features sensitive to variation in the spike rate.

Notably, at the high noise level $\beta = 0.8$, $\mathrm{MI}_F$ showed higher capture rate than the capture rate of the driving feature.
This is consistent in both processes, with a more marked difference observed for the AR(3) process: the capture rate of $\mathrm{MI}_F$ here is 52\%, whereas that of \verb|AC1| is only 14\%.
This improvement can be attributed to an ensemble effect, where the inclusion of multiple features in detecting dependence allows $\mathrm{MI}_F$ to successfully capture dependence if \textit{any} feature within the set manages to do so.
This ensemble approach proves highly beneficial in a high-noise regime where measurement noise significantly obscures information from the driving feature.
In other words, on occasions where $\mathrm{MI}_f$ corresponding to the true driving feature did not capture the interaction (here due to noise), other $\mathrm{MI}_f$ measures corresponding to correlated features may nevertheless stochastically capture the interaction and thereby enable a higher capture rate by $\mathrm{MI}_F$.

Overall, our results demonstrate that $\mathrm{MI}_F$ shows better performance in detecting dependencies than $\mathrm{MI}_s$ at all noise levels examined here for the two studied non-stationary processes, consistent our hypothesis that $\mathrm{MI}_F$ is more robust to noise when having relevant features in our capturing feature set.

\subsubsection{Detecting long-timescale interactions}
\label{sec:longtimescale}

\begin{figure*}[ht!]
    \centering
  \includegraphics[width=\textwidth]{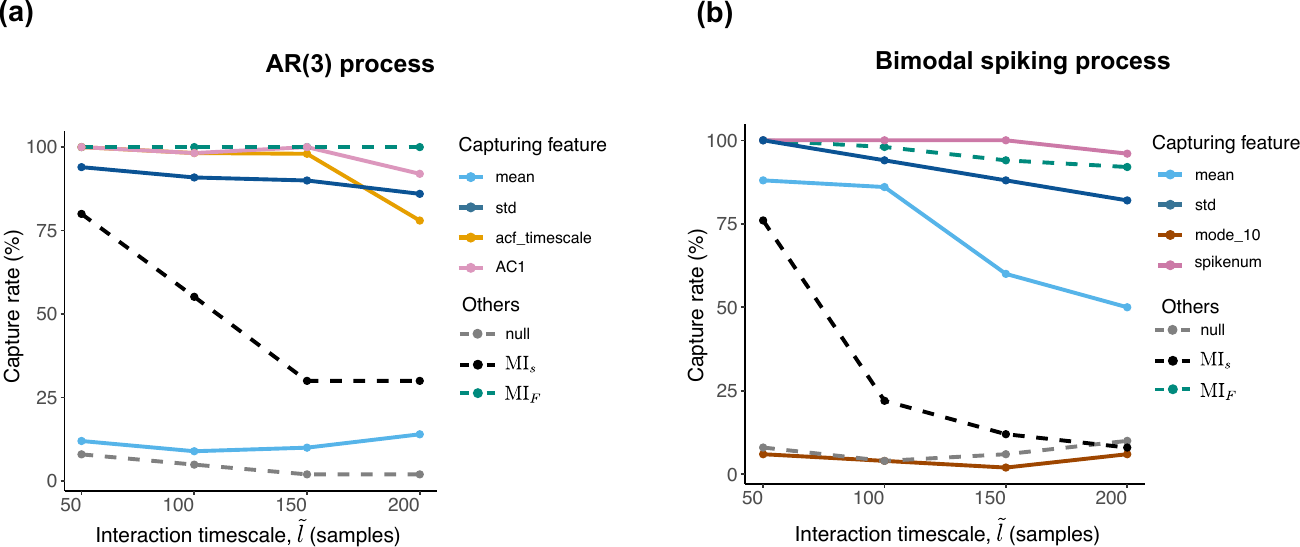}
  \caption{
  \textbf{$\mathrm{MI}_F$ exhibits improved detection of feature-mediated source--target interactions than $\mathrm{MI}_s$ for longer timescales of interaction.}
  We evaluated the capture rate of $\mathrm{MI}_F$ and $\mathrm{MI}_s$ across a range of interaction timescales, $\tilde{l} = 50, 100, 150, 200$ samples, for \textbf{(a)} a non-stationary AR(3) process (Eq.~\eqref{eqn:ar3}); and \textbf{(b)} a non-stationary bimodal spiking process [Eq.~\eqref{eqn:bimodal_spike_rate},\eqref{eqn:bimodal_spike_x}], for a time-series length $T = 1000$ samples and noise level $\beta = 0.6$.
  As indicated in the legends, we plot individual capturing features (solid lines), a comparison to the case for which the capturing feature is the driving feature (pink solid), $\mathrm{MI}_F$ using the full set of 24 candidate features (dark green dashed line), $\mathrm{MI}_s$ (black dashed line), and the noise time series $\mathbf{z}_\mathrm{null}$ (gray dashed).
  \label{fig:InteractionTimescaleAnalysis}  
  }
\end{figure*}

In Sec.~\ref{subsec:FIPI}, we highlighted that $\mathrm{MI}_s$ may face challenges when handling long-timescale interactions due to difficulties associated with estimating probability densities in a high-dimensional space.
We reasoned that, by transforming the source time series to features and inferring dependence from the feature space with lower dimensionality, $\mathrm{MI}_f$ can avoid the curse of dimensionality and thereby better detect long-timescale interactions.
Our objective here is to assess the performance of $\mathrm{MI}_F$ in comparison to $\mathrm{MI}_s$ in scenarios involving long interaction timescales.
We computed the capture rates of $\mathrm{MI}_F$ and $\mathrm{MI}_s$ for our two non-stationary processes---AR(3) and bimodal spiking---for time series of length $T = 1000$ samples across interaction timescales $\tilde{l} = 50, 100, 150$, and $200$ samples, while setting the noise level at $\beta = 0.6$.
We set this high noise level in order to study a challenging scenario, setting $\beta = 0.6$ based on our prior findings (see Fig.~\ref{fig:NoiseAnalysis}, where it is shown that this is the upper limit for which MI$_F$ exhibited capture rates over 90\% for both processes).

Results for the non-stationary AR(3) process and the bimodal spiking processes are shown in Figs~\ref{fig:InteractionTimescaleAnalysis}(a) and (b), respectively, illustrating the same features as those selected for presentation in Fig.~\ref{fig:NoiseAnalysis} above.
We found that $\mathrm{MI}_F$ outperforms $\mathrm{MI}_s$ across the full range of interaction timescales for both processes.
For the AR(3) process, $\mathrm{MI}_F$ showed 100\% capture rate across the full range of interaction timescales tested.
Top-performing features \verb|acf_timescale| and \verb|std| achieved capture rates $\geq 90\%$ at timescales up to $\tilde{l} = 150$ samples and over 75\% at $\tilde{l} = 200$ samples.
While $\mathrm{MI}_s$ performed well at $\tilde{l} = 50$, an already high-dimensional setting (with a high capture rate, 80\%), its capture rate dropped substantially with increasing $\tilde{l}$ (and hence dimensionality of the $\mathrm{MI}_s$ inference problem), dropping to just 30\% at $\tilde{l} = 150$ samples.

For the bimodal spiking process, shown in Fig.~\ref{fig:InteractionTimescaleAnalysis}(b), $\mathrm{MI}_F$ showed strong performance, with capture rates above 90\% for all interaction timescales tested.
Among the features, standard deviation (\verb|std|) exhibited strong performance, capturing a dependence with nearly 100\% capture rate at interaction timescales up to $\tilde{l} = 100$ samples, and over 75\% at timescales $\tilde{l} = 150$ and 200 samples.
Again, the performance of $\mathrm{MI}_s$ was overall inferior to $\mathrm{MI}_F$ across all tested interaction timescales, with its capture rate decreasingly steadily with increasing $\tilde{l}$.
These results indicate that $\mathrm{MI}_F$ can perform well at detecting time-series dependence on long timescales, even in a high noise setting, consistent with its ability to reduce high-dimensional sets of time-series measurements down to informative low-dimensional summaries.

\subsubsection{Detecting the timescale of interaction}
\label{sec:timescaledetection}

\begin{figure*}[ht!]
    \centering
  \includegraphics[width=1.00\textwidth]{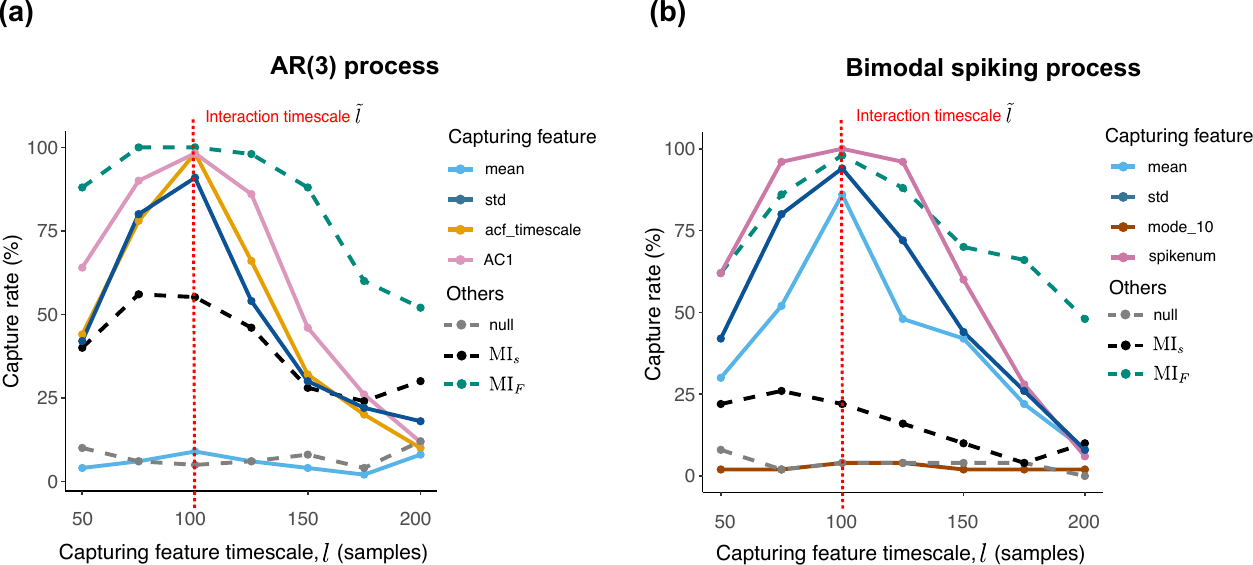}
  \caption{
  \textbf{The performance of $\mathrm{MI}_F$ peaks at the true timescale of interaction, $l = \tilde{l}$, suggesting an ability to infer the underlying timescale of source--target interaction.}
  We evaluated the performance of $\mathrm{MI}_F$ and $\mathrm{MI}_s$ across various capturing timescales $l$ for both \textbf{(a)} a non-stationary AR(3) process and \textbf{(b)} a non-stationary bimodal spiking process, each with a length of $T = 1000$ samples and noise level $\beta = 0.6$.
  Capture rates are plotted as a function of the capturing timescale within a which span a range from half to twice the interaction timescale: $0.5\tilde{l} \leq l \leq 2\tilde{l}$ ($l = 50, 75, 100, 125, 150, 175, 200$).  
  As indicated in the legends, we plot individual capturing features (solid lines), a comparison to the case for which the capturing feature is the driving feature computed over sliding windows of sized $l$ samples (pink solid), $\mathrm{MI}_F$ using the full set of 24 candidate features (dark green dashed line), $\mathrm{MI}_s$ (black dashed line), and the noise time series $\mathbf{z}_\mathrm{null}$ (gray dashed).
   An annotated vertical red line indicates the true interaction timescale $\tilde{l}$.
  }
  \label{fig:CapturingTimescaleAnalysis}
\end{figure*}

In the previous experiments, we provided $\mathrm{MI}_F$ and $\mathrm{MI}_s$ with the best chance to capture a source--target dependence by setting the window length of the capturing feature to align with the interaction timescale: $l = \tilde{l}$.
However, since in real-world scenarios the true interaction timescale is typically unknown, we next aimed to investigate how well we could capture dependence (and infer a true underlying timescale of interaction) in this more general case.
In the following experiment, we tested $\mathrm{MI}_F$ and $\mathrm{MI}_s$ on the non-stationary AR(3) and the bimodal spiking processes analyzed above, employing the same main parameter settings ($\beta = 0.6$ and $T = 1000$ samples) while removing the knowledge of the timescale of interaction $\tilde{l}$.
The interaction timescale was set to $\tilde{l} = 100$ samples for both processes.
In this setting, we compared the capture rate across a range of candidate timescales, which we set to be within the range of 50\% to 200\% of the interaction timescale $\tilde{l}$ ($l = 50, 75, 100, 125, 150, 175, 200$).
We expected to obtain a maximal capture rate when $l = \tilde{l}$ and aimed to understand how $\mathrm{MI}_F$ and $\mathrm{MI}_s$ perform when the true and tested timescales did not match, $l \neq \tilde{l}$.

The results are shown in Figs~\ref{fig:CapturingTimescaleAnalysis}(a) and (b) for the non-stationary AR(3) and the bimodal spiking processes, respectively.
For the AR(3) process, we found that $\mathrm{MI}_F$ could infer dependence reasonably well, with a capture rate above 85\% across the tested timescales from 50 to 150.
This suggests a high resilience to detecting a pairwise interaction even when using an inaccurate timescale (up to $|l-\tilde{l}| \approx 50$ samples), or 50\% of the interaction timescale.
For the bimodal spiking process, $\mathrm{MI}_F$ was resilient to detecting a pairwise interaction across a comparatively smaller range of candidate capturing timescales we tested; its capture rate is above 85\% for $75 \leq l \leq 125$, equivalent to an inaccuracy in timescale of up to 25\% of the interaction timescale [see Fig.~\ref{fig:CapturingTimescaleAnalysis}(b)].
The results from both processes indicates that $\mathrm{MI}_F$ can detect dependence even when the tested timescale of interaction is inaccurate, which is useful in practical scenarios where one has some hypothesis on the range of interaction timescales, without knowing the exact value.
Additionally, when the discrepancies between $l$ and $\tilde{l}$ are higher (e.g., for $l = 150, 175, 200$), $\mathrm{MI}_F$ outperformed $\mathrm{MI}_f$ when using the driving feature to infer dependence in both processes.
At first glance this may seem unexpected, but can be attributed to the ensemble effect of $\mathrm{MI}_F$ as  discussed in Sec.~\ref{sec:noise_robustness}, and the fact that the driving features used as capturing features here were computed over sliding window of size $l$ rather than $\tilde{l}$, introducing slight differences compared to the true driving features.

We also observed that, for both processes, the performance of $\mathrm{MI}_F$ peaked at the true interaction timescale, i.e., when $l = \tilde{l}$.
While $\mathrm{MI}_F$ is designed to infer dependence from a pair of time series, its application extends to scenarios where multiple realizations of these source and target time series exist.
An example of this is prevalent in neuroscience experiments, where neural activities from a pair of neurons are recorded across multiple subjects, each providing a distinct realization of the time series for the given pair of neurons.
The presence of multiple realizations for a pair of time-series variables allows us to calculate the percentage of dependence inferred at each candidate capturing timescale, where a peaked shape observed in the distribution of dependence percentages across timescales provides a valuable cue for identifying the true timescale of interaction.
By examining features with a peaked shape for the capture rate across the capturing timescales, our analysis approach additionally allows inference of the types of statistical properties that mediate the interaction.
In contrast, when using $\mathrm{MI}_s$ to infer interactions at different timescales, its capture rate fluctuates across different tested timescales and is generally low, indicating its limited ability to accurately infer the underlying interaction timescale. Specifically, across all tested timescales for both processes, the capture rate of $\mathrm{MI}_s$ was around 40\% for the non-stationary AR(3) process and about 16\% for the bimodal spiking process.

In summary, our comparisons of $\mathrm{MI}_F$ and $\mathrm{MI}_s$ across different noise levels, interaction timescales, and interaction timescales demonstrated the ability of $\mathrm{MI}_F$ to effectively capture longer timescale dependencies ($\tilde{l} > 100$), even when the timescale is not precisely set, as well as its robustness to higher noise levels than $\mathrm{MI}_s$.

\section{Discussion}
\label{sec:discussion}

This study introduces an information-theoretic approach to detect feature-mediated pairwise interactions between time series and analyzed its performance across a range of simulated settings.
We focused on comparing a mutual-information-based implementation of the approach, using a set of 24 features as $\mathrm{MI}_F$, to a more conventional formulation of mutual information, formulated in terms of the time-series values themselves, $\mathrm{MI}_s$.
We studied cases in which the source--target interaction is mediated by a statistical property of the source across a recent time period, corresponding to the case where a target is responding to non-stationary variation in the source that manifests as a change in the recent statistical properties of the source time series.
Our experiments confirm that this long-timescale setting is challenging for a conventional estimator like $\mathrm{MI}_s$, as it involves detecting a probabilistic dependence in a challenging high-dimensional space of recent time-series values.
By contrast, $\mathrm{MI}_F$ is more efficient in cases where it contains candidate features that are sensitive to the underlying interaction, due to its ability to distill complex time-series patterns down to a reduced statistical summary and thus perform the inference in a fixed, low-dimensional space, independent of the tested interaction timescale.
$\mathrm{MI}_F$ exhibits a greater performance in capturing feature-mediated dependencies, particularly for shorter time series (e.g., as low as 50 samples), longer timescales of dependence, and at higher noise levels.
Additionally, $\mathrm{MI}_F$ can detect source--target dependencies across a wider range of timescales around the true interaction timescale than MI$_s$.

A promising aspect of $\mathrm{MI}_F$ is that it allows researchers to select a list of features and timescales to test for dependence.
Our results demonstrate that other features that are sensitive to the driving feature can effectively detect a dependence mediated by the driving feature (even when this connection between these other features and the driving features is not obvious), and the same applies to timescales: features computed at timescales close to the ground-truth interaction timescale can capture the dependence almost as well as those computed at the true timescales in several cases.
Time-series features not only provide a more compact representation of the data, but also aid in comprehending the nature of dependence: they allow us to narrow down dynamical patterns between processes by selecting a specific set of features and examining potential dependencies between these features and the target process.
This flexibility in choosing the feature set and the intepretability of the features can provide valuable insights into the underlying dynamical interactions.
Researchers can use general-purpose feature sets \cite{Fulcher2013hctsa, feasts_pkg, Lubba2019Catch22} and can also choose to iterate through a range of timescales to infer the underlying dependence structure, making $\mathrm{MI}_F$ a powerful tool for data-driven analyses.
Leveraging the feature-based approach effectively reduces the dimensionality of the learning space of the information-theoretic estimator, leading to faster computation and more efficient representation of the relevant joint probability of the source and target processes.
This enables researchers to iterate through different feature windows and uncover dependencies between the source and target processes more efficiently with respect to both computation time and data availability.

Interestingly, our results highlight that by using an ensemble of features including those that are sensitive to the driving feature, $\mathrm{MI}_F$ can exhibit better performance than using the feature directly mediating the interaction as capturing feature, in scenarios with high levels of noise or significant deviations in the capturing timescale from the actual interaction timescale.
This phenomenon can be explained by the concepts of information synergy and redundancy \cite{Griffith2014InfoSynergy, Timme2014SynergyRedundancy}, where combining information from multiple features surpasses the sum of individual contributions (information synergy) or exploits overlapping information across features (information redundancy).
Information synergy has been shown to be associated with resilience to noise \cite{Quax2017QuantifySynergisticInfo}, while some studies have shown that feature space, as a redundant representation of the signal, may have noise reduction properties that help with reconstructing the signal \cite{Rozell2006RedundantPopulationCode, Rozell2005RedundancyNoiseReduction} (the driving feature can be considered as the `signal' which needs to be reconstructed in this context).  
This theoretical foundation underscores the promise of employing a diverse set of features in the inference of feature-mediated dependence.

However, this feature-based approach also presents challenges.
Transforming time-series data into a set of time-series features results in the loss of some information from the original data, and if the candidate feature set $F$ does not contain any features that are sensitive to the dynamical property mediating the source--target interaction, $\mathrm{MI}_F$ will not be able to detect the dependence.
Although this could be countered by using a large set of candidate time-series features in the absence of strong domain or prior knowledge on the potential driving features and timescales of interactions \cite{Fulcher2017:HctsaComputationalFramework}, comparing the behavior of a potentially large pool of candidate features and timescales can lead to a high multiple-hypothesis burden and potentially yielding false negatives.
This process of selecting a concise yet appropriately expressive feature set can thus be important and should be informed by relevant hypotheses, dataset size, expected effect size, and other relevant factors where possible.
Furthermore, as discussed previously, the presence of redundant information in a non-orthogonal feature space can prove beneficial in high-noise regimes.
Therefore having some correlated features in the capturing feature set should be considered, especially when dealing with exceptionally noisy systems.

While studying the performance of the $\mathrm{MI}_F$ and $\mathrm{MI}_s$ across different generative processes, we found that the detection power of $\mathrm{MI}_s$ depends on multiple factors: the timescale of the interaction, the features being used to generate the target, and the characteristics of the source processes.
The interplay of these properties results in a non-monotonic relationship between the dependence capture rate and any single factor.
For example, from our experiment with random noise process we can see that capture rate for dependence mediated by distribution mode (\verb|mode_5|) has much slower improvement with increasing time-series length than the mean (\verb|mean|), autocorrelation timescale (\verb|acf_timescale|), and a time-reversibility metric (\verb|trev|).
This implies that interactions mediated by some features of the dynamics are harder to detect than those mediated by others. 
As the features of a time series are mathematical transformations of its elements, $z_t = f(\mathbf{x}_t^{(l)})$, we hypothesize that this difference could be driven by different levels of roughness of the feature surface in the time-delay embedding space $\mathbb{R}^{l\times {(T-l+1)}}$.
That is, we expect a relatively smooth surface to be less vulnerable to noise and thus require fewer data points for $\mathrm{MI}_s$ to reliably estimate the required probability densities.
For example, \verb|mean|, \verb|std| and \verb|trev| are computed as polynomials of degree up to three in the time-delay embedding space of the source process, their surfaces are relatively smooth compared to \verb|mode_5|, whose computation involves multiple nonlinear operations.
More research is needed to fully validate and refine this hypothesis, providing a clearer understanding of the underlying mechanisms that contribute to the observed performance differences in  $\mathrm{MI}_F$ and $\mathrm{MI}_s$ across diverse scenarios and features. 
This could help inform their strategic application for specific cases.

Delving into potential extensions of our method, we acknowledge that our focus here has been on the specific scenario of a target process responding to a single time-series feature from the recent past of the source [as depicted in Fig.~\ref{fig:methodschematic}(a)].
Many alternative settings are possible and would make straightforward extensions of the theory and algorithms introduced here.
One such direction involves extending the method to account for the case that the target process has non-trivial dynamics [extending the simple case of a linear noisy response used here, Eq.~\eqref{eqn:driving_target}], in which case using transfer entropy instead of mutual information could offer a more accurate detection and understanding of the underlying dependence structure.
In our simulation studies, we have also neglected a time-delay from source to target, considering $y_t$ to respond to a past window of $x$ from $t - \tau_1$ to $t$, which could be trivially generalized to a window of $x$ from $t - \tau_1$ to $t - \tau_2$.
Another more general setting that could be studied involves the target process responding to \textit{multiple} features of the source process (i.e., $\mathbf{z}$ becomes multivariate).
An even more challenging scenario has one or more \textit{features} of the target process (rather than values, $y_t$, as considered here) responding to one or more features of the source, such that the dependence is inferred from the feature vectors of both the source and target time series.
By computing and concatenating the features of interest as an embedding search space, we could then compute mutual information or transfer entropy over this embedded feature space and the target vector, revealing complex interactions between multiple source and target dynamical properties.

Combining the feature-based approach with the Information Bottleneck framework \cite{Tishby1999InformationBottleneck} presents a promising avenue for further research.
This hybrid approach offers a concrete method to construct a lower-dimensional representation of the data and verify whether there is dependence between two variables.
Subsequently, the Information Bottleneck framework can be used to identify the best representation that maximizes the mutual information between these variables, providing a refined understanding of the underlying dependence structure and finding a compact representation of the data while preserving its information.

In conclusion, our feature-based information-theoretic approach is shown to be effective in capturing long-timescale feature-mediated dependence from time series.
Its sensitivity, computational efficiency, and interpretability make it a valuable tool for a wide range of applications, especially in dealing with noisy time-series data.
We anticipate it to contribute to advancing our understanding of complex systems and their underlying dynamics in real-world datasets.

\begin{acknowledgments}
We thank Brendan Harris and Annie Bryant for computational assistance and helpful discussions.
\end{acknowledgments}

\bibliography{refs.bib}

\begin{thebibliography}{60}%
\makeatletter
\providecommand \@ifxundefined [1]{%
 \@ifx{#1\undefined}
}%
\providecommand \@ifnum [1]{%
 \ifnum #1\expandafter \@firstoftwo
 \else \expandafter \@secondoftwo
 \fi
}%
\providecommand \@ifx [1]{%
 \ifx #1\expandafter \@firstoftwo
 \else \expandafter \@secondoftwo
 \fi
}%
\providecommand \natexlab [1]{#1}%
\providecommand \enquote  [1]{``#1''}%
\providecommand \bibnamefont  [1]{#1}%
\providecommand \bibfnamefont [1]{#1}%
\providecommand \citenamefont [1]{#1}%
\providecommand \href@noop [0]{\@secondoftwo}%
\providecommand \href [0]{\begingroup \@sanitize@url \@href}%
\providecommand \@href[1]{\@@startlink{#1}\@@href}%
\providecommand \@@href[1]{\endgroup#1\@@endlink}%
\providecommand \@sanitize@url [0]{\catcode `\\12\catcode `\$12\catcode
  `\&12\catcode `\#12\catcode `\^12\catcode `\_12\catcode `\%12\relax}%
\providecommand \@@startlink[1]{}%
\providecommand \@@endlink[0]{}%
\providecommand \url  [0]{\begingroup\@sanitize@url \@url }%
\providecommand \@url [1]{\endgroup\@href {#1}{\urlprefix }}%
\providecommand \urlprefix  [0]{URL }%
\providecommand \Eprint [0]{\href }%
\providecommand \doibase [0]{https://doi.org/}%
\providecommand \selectlanguage [0]{\@gobble}%
\providecommand \bibinfo  [0]{\@secondoftwo}%
\providecommand \bibfield  [0]{\@secondoftwo}%
\providecommand \translation [1]{[#1]}%
\providecommand \BibitemOpen [0]{}%
\providecommand \bibitemStop [0]{}%
\providecommand \bibitemNoStop [0]{.\EOS\space}%
\providecommand \EOS [0]{\spacefactor3000\relax}%
\providecommand \BibitemShut  [1]{\csname bibitem#1\endcsname}%
\let\auto@bib@innerbib\@empty
\bibitem [{\citenamefont {Peel}\ \emph {et~al.}(2022)\citenamefont {Peel},
  \citenamefont {Peixoto},\ and\ \citenamefont
  {De~Domenico}}]{Peel2022StatisticalInferenceNetworkScience}%
  \BibitemOpen
  \bibfield  {author} {\bibinfo {author} {\bibfnamefont {L.}~\bibnamefont
  {Peel}}, \bibinfo {author} {\bibfnamefont {T.~P.}\ \bibnamefont {Peixoto}},\
  and\ \bibinfo {author} {\bibfnamefont {M.}~\bibnamefont {De~Domenico}},\
  }\bibfield  {title} {\bibinfo {title} {Statistical inference links data and
  theory in network science},\ }\href
  {https://doi.org/10.1038/s41467-022-34267-9} {\bibfield  {journal} {\bibinfo
  {journal} {Nature Communications}\ }\textbf {\bibinfo {volume} {13}},\
  \bibinfo {pages} {6794} (\bibinfo {year} {2022})}\BibitemShut {NoStop}%
\bibitem [{\citenamefont {Runge}\ \emph {et~al.}(2019)\citenamefont {Runge},
  \citenamefont {Bathiany}, \citenamefont {Bollt}, \citenamefont
  {{Camps-Valls}}, \citenamefont {Coumou}, \citenamefont {Deyle}, \citenamefont
  {Glymour}, \citenamefont {Kretschmer}, \citenamefont {Mahecha}, \citenamefont
  {{Mu{\~n}oz-Mar{\'i}}}, \citenamefont {{van Nes}}, \citenamefont {Peters},
  \citenamefont {Quax}, \citenamefont {Reichstein}, \citenamefont {Scheffer},
  \citenamefont {Sch{\"o}lkopf}, \citenamefont {Spirtes}, \citenamefont
  {Sugihara}, \citenamefont {Sun}, \citenamefont {Zhang},\ and\ \citenamefont
  {Zscheischler}}]{Runge2019CausalInferenceEarthScience}%
  \BibitemOpen
  \bibfield  {author} {\bibinfo {author} {\bibfnamefont {J.}~\bibnamefont
  {Runge}}, \bibinfo {author} {\bibfnamefont {S.}~\bibnamefont {Bathiany}},
  \bibinfo {author} {\bibfnamefont {E.}~\bibnamefont {Bollt}}, \bibinfo
  {author} {\bibfnamefont {G.}~\bibnamefont {{Camps-Valls}}}, \bibinfo {author}
  {\bibfnamefont {D.}~\bibnamefont {Coumou}}, \bibinfo {author} {\bibfnamefont
  {E.}~\bibnamefont {Deyle}}, \bibinfo {author} {\bibfnamefont
  {C.}~\bibnamefont {Glymour}}, \bibinfo {author} {\bibfnamefont
  {M.}~\bibnamefont {Kretschmer}}, \bibinfo {author} {\bibfnamefont {M.~D.}\
  \bibnamefont {Mahecha}}, \bibinfo {author} {\bibfnamefont {J.}~\bibnamefont
  {{Mu{\~n}oz-Mar{\'i}}}}, \bibinfo {author} {\bibfnamefont {E.~H.}\
  \bibnamefont {{van Nes}}}, \bibinfo {author} {\bibfnamefont {J.}~\bibnamefont
  {Peters}}, \bibinfo {author} {\bibfnamefont {R.}~\bibnamefont {Quax}},
  \bibinfo {author} {\bibfnamefont {M.}~\bibnamefont {Reichstein}}, \bibinfo
  {author} {\bibfnamefont {M.}~\bibnamefont {Scheffer}}, \bibinfo {author}
  {\bibfnamefont {B.}~\bibnamefont {Sch{\"o}lkopf}}, \bibinfo {author}
  {\bibfnamefont {P.}~\bibnamefont {Spirtes}}, \bibinfo {author} {\bibfnamefont
  {G.}~\bibnamefont {Sugihara}}, \bibinfo {author} {\bibfnamefont
  {J.}~\bibnamefont {Sun}}, \bibinfo {author} {\bibfnamefont {K.}~\bibnamefont
  {Zhang}},\ and\ \bibinfo {author} {\bibfnamefont {J.}~\bibnamefont
  {Zscheischler}},\ }\bibfield  {title} {\bibinfo {title} {Inferring causation
  from time series in {{Earth}} system sciences},\ }\href
  {https://doi.org/10.1038/s41467-019-10105-3} {\bibfield  {journal} {\bibinfo
  {journal} {Nature Communications}\ }\textbf {\bibinfo {volume} {10}},\
  \bibinfo {pages} {2553} (\bibinfo {year} {2019})}\BibitemShut {NoStop}%
\bibitem [{\citenamefont {Hoffmann}\ \emph {et~al.}(2020)\citenamefont
  {Hoffmann}, \citenamefont {Peel}, \citenamefont {Lambiotte},\ and\
  \citenamefont {{Nick S. Jones}}}]{Hoffmann2020ComunityDetection}%
  \BibitemOpen
  \bibfield  {author} {\bibinfo {author} {\bibfnamefont {T.}~\bibnamefont
  {Hoffmann}}, \bibinfo {author} {\bibfnamefont {L.}~\bibnamefont {Peel}},
  \bibinfo {author} {\bibfnamefont {R.}~\bibnamefont {Lambiotte}},\ and\
  \bibinfo {author} {\bibnamefont {{Nick S. Jones}}},\ }\bibfield  {title}
  {\bibinfo {title} {Community detection in networks without observing edges},\
  }\href {https://doi.org/10.1126/sciadv.aav1478} {\bibfield  {journal}
  {\bibinfo  {journal} {Science Advances}\ }\textbf {\bibinfo {volume} {6}},\
  \bibinfo {pages} {eaav1478} (\bibinfo {year} {2020})}\BibitemShut {NoStop}%
\bibitem [{\citenamefont {Lef{\`e}vre}\ \emph {et~al.}(2014)\citenamefont
  {Lef{\`e}vre}, \citenamefont {Vasquez},\ and\ \citenamefont
  {Laugier}}]{Lefevre2014MotionPrediction}%
  \BibitemOpen
  \bibfield  {author} {\bibinfo {author} {\bibfnamefont {S.}~\bibnamefont
  {Lef{\`e}vre}}, \bibinfo {author} {\bibfnamefont {D.}~\bibnamefont
  {Vasquez}},\ and\ \bibinfo {author} {\bibfnamefont {C.}~\bibnamefont
  {Laugier}},\ }\bibfield  {title} {\bibinfo {title} {A survey on motion
  prediction and risk assessment for intelligent vehicles},\ }\href
  {https://doi.org/10.1186/s40648-014-0001-z} {\bibfield  {journal} {\bibinfo
  {journal} {ROBOMECH Journal}\ }\textbf {\bibinfo {volume} {1}},\ \bibinfo
  {pages} {1} (\bibinfo {year} {2014})}\BibitemShut {NoStop}%
\bibitem [{\citenamefont {Freedman}\ \emph {et~al.}(2007)\citenamefont
  {Freedman}, \citenamefont {Pisani},\ and\ \citenamefont
  {Purves}}]{freedman2007statistics}%
  \BibitemOpen
  \bibfield  {author} {\bibinfo {author} {\bibfnamefont {D.}~\bibnamefont
  {Freedman}}, \bibinfo {author} {\bibfnamefont {R.}~\bibnamefont {Pisani}},\
  and\ \bibinfo {author} {\bibfnamefont {R.}~\bibnamefont {Purves}},\
  }\bibfield  {title} {\bibinfo {title} {Statistics (international student
  edition)},\ }\href@noop {} {\bibfield  {journal} {\bibinfo  {journal}
  {Pisani, R. Purves, 4th edn. WW Norton \& Company, New York}\ } (\bibinfo
  {year} {2007})}\BibitemShut {NoStop}%
\bibitem [{\citenamefont {Granger}(2001)}]{Granger2001causality}%
  \BibitemOpen
  \bibfield  {author} {\bibinfo {author} {\bibfnamefont {C.~W.~J.}\
  \bibnamefont {Granger}},\ }\bibfield  {title} {\bibinfo {title}
  {Investigating causal relations by econometric models and cross-spectral
  methods},\ }in\ \href@noop {} {\emph {\bibinfo {booktitle} {Essays in
  Econometrics: {{Collected}} Papers of Clive {{W}}. {{J}}. {{Granger}}}}}\
  (\bibinfo  {publisher} {{Harvard University Press}},\ \bibinfo {address}
  {{USA}},\ \bibinfo {year} {2001})\ pp.\ \bibinfo {pages} {31--47}\BibitemShut
  {NoStop}%
\bibitem [{\citenamefont {Cliff}\ \emph {et~al.}(2023)\citenamefont {Cliff},
  \citenamefont {Bryant}, \citenamefont {Lizier}, \citenamefont {Tsuchiya},\
  and\ \citenamefont {Fulcher}}]{Cliff2023UnifyingPairwiseInteractions}%
  \BibitemOpen
  \bibfield  {author} {\bibinfo {author} {\bibfnamefont {O.~M.}\ \bibnamefont
  {Cliff}}, \bibinfo {author} {\bibfnamefont {A.~G.}\ \bibnamefont {Bryant}},
  \bibinfo {author} {\bibfnamefont {J.~T.}\ \bibnamefont {Lizier}}, \bibinfo
  {author} {\bibfnamefont {N.}~\bibnamefont {Tsuchiya}},\ and\ \bibinfo
  {author} {\bibfnamefont {B.~D.}\ \bibnamefont {Fulcher}},\ }\bibfield
  {title} {\bibinfo {title} {Unifying pairwise interactions in complex
  dynamics},\ }\href {https://doi.org/10.1038/s43588-023-00519-x} {\bibfield
  {journal} {\bibinfo  {journal} {Nature Computational Science}\ }\textbf
  {\bibinfo {volume} {3}},\ \bibinfo {pages} {883} (\bibinfo {year}
  {2023})}\BibitemShut {NoStop}%
\bibitem [{\citenamefont {MacKay}(2003)}]{MacKay2003informationtheory}%
  \BibitemOpen
  \bibfield  {author} {\bibinfo {author} {\bibfnamefont {D.~J.~C.}\
  \bibnamefont {MacKay}},\ }\href@noop {} {\emph {\bibinfo {title} {Information
  Theory, Inference, and Learning Algorithms}}}\ (\bibinfo  {publisher}
  {{Cambridge University Press}},\ \bibinfo {year} {2003})\BibitemShut
  {NoStop}%
\bibitem [{\citenamefont {Schreiber}(2000)}]{Schreiber2000TE}%
  \BibitemOpen
  \bibfield  {author} {\bibinfo {author} {\bibfnamefont {T.}~\bibnamefont
  {Schreiber}},\ }\bibfield  {title} {\bibinfo {title} {Measuring information
  transfer},\ }\href {https://doi.org/10.1103/physrevlett.85.461} {\bibfield
  {journal} {\bibinfo  {journal} {Physical Review Letters}\ }\textbf {\bibinfo
  {volume} {85}},\ \bibinfo {pages} {461} (\bibinfo {year} {2000})}\BibitemShut
  {NoStop}%
\bibitem [{\citenamefont {Bossomaier}\ \emph {et~al.}(2016)\citenamefont
  {Bossomaier}, \citenamefont {Barnett}, \citenamefont {Harr{\'e}},\ and\
  \citenamefont {Lizier}}]{Bossomaier2016TEIntro}%
  \BibitemOpen
  \bibfield  {author} {\bibinfo {author} {\bibfnamefont {T.~R.~J.}\
  \bibnamefont {Bossomaier}}, \bibinfo {author} {\bibfnamefont {L.~C.}\
  \bibnamefont {Barnett}}, \bibinfo {author} {\bibfnamefont {M.}~\bibnamefont
  {Harr{\'e}}},\ and\ \bibinfo {author} {\bibfnamefont {J.~T.}\ \bibnamefont
  {Lizier}},\ }\href {https://doi.org/10.1007/978-3-319-43222-9} {\emph
  {\bibinfo {title} {An Introduction to Transfer Entropy: {{Information}} Flow
  in Complex Systems}}}\ (\bibinfo  {publisher} {{Springer}},\ \bibinfo
  {address} {{Cham}},\ \bibinfo {year} {2016})\BibitemShut {NoStop}%
\bibitem [{\citenamefont {Materassi}\ \emph {et~al.}(2014)\citenamefont
  {Materassi}, \citenamefont {Consolini}, \citenamefont {Smith},\ and\
  \citenamefont {De~Marco}}]{Materassi2014TEFluidTurbulence}%
  \BibitemOpen
  \bibfield  {author} {\bibinfo {author} {\bibfnamefont {M.}~\bibnamefont
  {Materassi}}, \bibinfo {author} {\bibfnamefont {G.}~\bibnamefont
  {Consolini}}, \bibinfo {author} {\bibfnamefont {N.}~\bibnamefont {Smith}},\
  and\ \bibinfo {author} {\bibfnamefont {R.}~\bibnamefont {De~Marco}},\
  }\bibfield  {title} {\bibinfo {title} {Information theory analysis of
  cascading process in a synthetic model of fluid turbulence},\ }\href
  {https://doi.org/10.3390/e16031272} {\bibfield  {journal} {\bibinfo
  {journal} {Entropy}\ }\textbf {\bibinfo {volume} {16}},\ \bibinfo {pages}
  {1272} (\bibinfo {year} {2014})}\BibitemShut {NoStop}%
\bibitem [{\citenamefont {Ursino}\ \emph {et~al.}(2020)\citenamefont {Ursino},
  \citenamefont {Ricci},\ and\ \citenamefont
  {Magosso}}]{Ursino2020TEBrainConnectivity}%
  \BibitemOpen
  \bibfield  {author} {\bibinfo {author} {\bibfnamefont {M.}~\bibnamefont
  {Ursino}}, \bibinfo {author} {\bibfnamefont {G.}~\bibnamefont {Ricci}},\ and\
  \bibinfo {author} {\bibfnamefont {E.}~\bibnamefont {Magosso}},\ }\bibfield
  {title} {\bibinfo {title} {Transfer entropy as a measure of brain
  connectivity: {{A}} critical analysis with the help of neural mass models},\
  }\href {https://doi.org/10.3389/fncom.2020.00045} {\bibfield  {journal}
  {\bibinfo  {journal} {Frontiers in Computational Neuroscience}\ }\textbf
  {\bibinfo {volume} {14}},\ \bibinfo {pages} {45} (\bibinfo {year}
  {2020})}\BibitemShut {NoStop}%
\bibitem [{\citenamefont {Dimpfl}\ and\ \citenamefont
  {Peter}(2013)}]{Dimpfl2013TEInformationFlowFinancec}%
  \BibitemOpen
  \bibfield  {author} {\bibinfo {author} {\bibfnamefont {T.}~\bibnamefont
  {Dimpfl}}\ and\ \bibinfo {author} {\bibfnamefont {F.~J.}\ \bibnamefont
  {Peter}},\ }\bibfield  {title} {\bibinfo {title} {Using transfer entropy to
  measure information flows between financial markets},\ }\href
  {https://doi.org/doi:10.1515/snde-2012-0044} {\bibfield  {journal} {\bibinfo
  {journal} {Studies in Nonlinear Dynamics and Econometrics}\ }\textbf
  {\bibinfo {volume} {17}},\ \bibinfo {pages} {85} (\bibinfo {year}
  {2013})}\BibitemShut {NoStop}%
\bibitem [{\citenamefont {Li}\ \emph {et~al.}(2013)\citenamefont {Li},
  \citenamefont {Liang}, \citenamefont {Zhu}, \citenamefont {Sun},\ and\
  \citenamefont {Wu}}]{Li2013TERiskContagion}%
  \BibitemOpen
  \bibfield  {author} {\bibinfo {author} {\bibfnamefont {J.}~\bibnamefont
  {Li}}, \bibinfo {author} {\bibfnamefont {C.}~\bibnamefont {Liang}}, \bibinfo
  {author} {\bibfnamefont {X.}~\bibnamefont {Zhu}}, \bibinfo {author}
  {\bibfnamefont {X.}~\bibnamefont {Sun}},\ and\ \bibinfo {author}
  {\bibfnamefont {D.}~\bibnamefont {Wu}},\ }\bibfield  {title} {\bibinfo
  {title} {Risk contagion in chinese banking industry: {{A}} transfer
  entropy-based analysis},\ }\href {https://doi.org/10.3390/e15125549}
  {\bibfield  {journal} {\bibinfo  {journal} {Entropy}\ }\textbf {\bibinfo
  {volume} {15}},\ \bibinfo {pages} {5549} (\bibinfo {year}
  {2013})}\BibitemShut {NoStop}%
\bibitem [{\citenamefont {Li}\ and\ \citenamefont
  {Giles}(2015)}]{Li2015StockVolatilitySpillover}%
  \BibitemOpen
  \bibfield  {author} {\bibinfo {author} {\bibfnamefont {Y.}~\bibnamefont
  {Li}}\ and\ \bibinfo {author} {\bibfnamefont {D.~E.}\ \bibnamefont {Giles}},\
  }\bibfield  {title} {\bibinfo {title} {Modelling volatility spillover effects
  between developed stock markets and asian emerging stock markets},\ }\href
  {https://doi.org/10.1002/ijfe.1506} {\bibfield  {journal} {\bibinfo
  {journal} {International Journal of Finance \& Economics}\ }\textbf {\bibinfo
  {volume} {20}},\ \bibinfo {pages} {155} (\bibinfo {year} {2015})}\BibitemShut
  {NoStop}%
\bibitem [{\citenamefont {Bonnefond}\ \emph {et~al.}(2017)\citenamefont
  {Bonnefond}, \citenamefont {Kastner},\ and\ \citenamefont
  {Jensen}}]{Bonnefond2017OscillationBasedBrainCommunication}%
  \BibitemOpen
  \bibfield  {author} {\bibinfo {author} {\bibfnamefont {M.}~\bibnamefont
  {Bonnefond}}, \bibinfo {author} {\bibfnamefont {S.}~\bibnamefont {Kastner}},\
  and\ \bibinfo {author} {\bibfnamefont {O.}~\bibnamefont {Jensen}},\
  }\bibfield  {title} {\bibinfo {title} {Communication between brain areas
  based on nested oscillations},\ }\href
  {https://doi.org/10.1523/ENEURO.0153-16.2017} {\bibfield  {journal} {\bibinfo
   {journal} {eNeuro}\ }\textbf {\bibinfo {volume} {4}},\ \bibinfo {pages}
  {ENEURO.0153} (\bibinfo {year} {2017})}\BibitemShut {NoStop}%
\bibitem [{\citenamefont {Markicevic}\ \emph {et~al.}(2020)\citenamefont
  {Markicevic}, \citenamefont {Fulcher}, \citenamefont {Lewis}, \citenamefont
  {Helmchen}, \citenamefont {Rudin}, \citenamefont {Zerbi},\ and\ \citenamefont
  {Wenderoth}}]{Markicevic2020EIBrainDisorders}%
  \BibitemOpen
  \bibfield  {author} {\bibinfo {author} {\bibfnamefont {M.}~\bibnamefont
  {Markicevic}}, \bibinfo {author} {\bibfnamefont {B.~D.}\ \bibnamefont
  {Fulcher}}, \bibinfo {author} {\bibfnamefont {C.}~\bibnamefont {Lewis}},
  \bibinfo {author} {\bibfnamefont {F.}~\bibnamefont {Helmchen}}, \bibinfo
  {author} {\bibfnamefont {M.}~\bibnamefont {Rudin}}, \bibinfo {author}
  {\bibfnamefont {V.}~\bibnamefont {Zerbi}},\ and\ \bibinfo {author}
  {\bibfnamefont {N.}~\bibnamefont {Wenderoth}},\ }\bibfield  {title} {\bibinfo
  {title} {Cortical {{Excitation}}:{{Inhibition}} imbalance causes abnormal
  brain network dynamics as observed in neurodevelopmental disorders},\ }\href
  {https://doi.org/10.1093/cercor/bhaa084} {\bibfield  {journal} {\bibinfo
  {journal} {Cerebral Cortex}\ }\textbf {\bibinfo {volume} {30}},\ \bibinfo
  {pages} {4922} (\bibinfo {year} {2020})}\BibitemShut {NoStop}%
\bibitem [{\citenamefont {Trakoshis}\ \emph {et~al.}(2020)\citenamefont
  {Trakoshis}, \citenamefont {{Mart{\'i}nez-Ca{\~n}ada}}, \citenamefont
  {Rocchi}, \citenamefont {Canella}, \citenamefont {You}, \citenamefont
  {Chakrabarti}, \citenamefont {Ruigrok}, \citenamefont {Bullmore},
  \citenamefont {Suckling}, \citenamefont {Markicevic}, \citenamefont {Zerbi},
  \citenamefont {Consortium}, \citenamefont {{Baron-Cohen}}, \citenamefont
  {Gozzi}, \citenamefont {Lai}, \citenamefont {Panzeri}, \citenamefont
  {Lombardo}, \citenamefont {B{\"u}chel}, \citenamefont {Breakspear},
  \citenamefont {Watanabe},\ and\ \citenamefont {Gao}}]{Trakoshis2020EI}%
  \BibitemOpen
  \bibfield  {author} {\bibinfo {author} {\bibfnamefont {S.}~\bibnamefont
  {Trakoshis}}, \bibinfo {author} {\bibfnamefont {P.}~\bibnamefont
  {{Mart{\'i}nez-Ca{\~n}ada}}}, \bibinfo {author} {\bibfnamefont
  {F.}~\bibnamefont {Rocchi}}, \bibinfo {author} {\bibfnamefont
  {C.}~\bibnamefont {Canella}}, \bibinfo {author} {\bibfnamefont
  {W.}~\bibnamefont {You}}, \bibinfo {author} {\bibfnamefont {B.}~\bibnamefont
  {Chakrabarti}}, \bibinfo {author} {\bibfnamefont {A.~N.~V.}\ \bibnamefont
  {Ruigrok}}, \bibinfo {author} {\bibfnamefont {E.~T.}\ \bibnamefont
  {Bullmore}}, \bibinfo {author} {\bibfnamefont {J.}~\bibnamefont {Suckling}},
  \bibinfo {author} {\bibfnamefont {M.}~\bibnamefont {Markicevic}}, \bibinfo
  {author} {\bibfnamefont {V.}~\bibnamefont {Zerbi}}, \bibinfo {author}
  {\bibfnamefont {M.~R. C. A. I. M.~S.}\ \bibnamefont {Consortium}}, \bibinfo
  {author} {\bibfnamefont {S.}~\bibnamefont {{Baron-Cohen}}}, \bibinfo {author}
  {\bibfnamefont {A.}~\bibnamefont {Gozzi}}, \bibinfo {author} {\bibfnamefont
  {M.-C.}\ \bibnamefont {Lai}}, \bibinfo {author} {\bibfnamefont
  {S.}~\bibnamefont {Panzeri}}, \bibinfo {author} {\bibfnamefont {M.~V.}\
  \bibnamefont {Lombardo}}, \bibinfo {author} {\bibfnamefont {C.}~\bibnamefont
  {B{\"u}chel}}, \bibinfo {author} {\bibfnamefont {M.}~\bibnamefont
  {Breakspear}}, \bibinfo {author} {\bibfnamefont {T.}~\bibnamefont
  {Watanabe}},\ and\ \bibinfo {author} {\bibfnamefont {R.}~\bibnamefont
  {Gao}},\ }\bibfield  {title} {\bibinfo {title} {Intrinsic
  excitation-inhibition imbalance affects medial prefrontal cortex differently
  in autistic men versus women},\ }\href {https://doi.org/10.7554/eLife.55684}
  {\bibfield  {journal} {\bibinfo  {journal} {eLife}\ }\textbf {\bibinfo
  {volume} {9}},\ \bibinfo {pages} {e55684} (\bibinfo {year}
  {2020})}\BibitemShut {NoStop}%
\bibitem [{\citenamefont {Albuquerque}\ \emph {et~al.}(2021)\citenamefont
  {Albuquerque}, \citenamefont {Valente}, \citenamefont {Teixeira},
  \citenamefont {Figueiredo}, \citenamefont {Sa-Couto},\ and\ \citenamefont
  {Oliveira}}]{Albuquerque2021SpeechAnxiety}%
  \BibitemOpen
  \bibfield  {author} {\bibinfo {author} {\bibfnamefont {L.}~\bibnamefont
  {Albuquerque}}, \bibinfo {author} {\bibfnamefont {A.~R.~S.}\ \bibnamefont
  {Valente}}, \bibinfo {author} {\bibfnamefont {A.}~\bibnamefont {Teixeira}},
  \bibinfo {author} {\bibfnamefont {D.}~\bibnamefont {Figueiredo}}, \bibinfo
  {author} {\bibfnamefont {P.}~\bibnamefont {Sa-Couto}},\ and\ \bibinfo
  {author} {\bibfnamefont {C.}~\bibnamefont {Oliveira}},\ }\bibfield  {title}
  {\bibinfo {title} {Association between acoustic speech features and
  non-severe levels of anxiety and depression symptoms across lifespan},\
  }\href {https://doi.org/10.1371/journal.pone.0248842} {\bibfield  {journal}
  {\bibinfo  {journal} {PLOS ONE}\ }\textbf {\bibinfo {volume} {16}},\ \bibinfo
  {pages} {1} (\bibinfo {year} {2021})}\BibitemShut {NoStop}%
\bibitem [{\citenamefont {Takens}(1981)}]{Taken1981TakenEmbeddings}%
  \BibitemOpen
  \bibfield  {author} {\bibinfo {author} {\bibfnamefont {F.}~\bibnamefont
  {Takens}},\ }\bibfield  {title} {\bibinfo {title} {Detecting strange
  attractors in turbulence},\ }in\ \href {https://doi.org/10.1007/bfb0091924}
  {\emph {\bibinfo {booktitle} {Dynamical Systems and Turbulence, Warwick
  1980}}},\ \bibinfo {series} {Lecture Notes in Mathematics}, Vol.\ \bibinfo
  {volume} {898},\ \bibinfo {editor} {edited by\ \bibinfo {editor}
  {\bibfnamefont {D.}~\bibnamefont {Rand}}\ and\ \bibinfo {editor}
  {\bibfnamefont {L.-S.}\ \bibnamefont {Young}}}\ (\bibinfo  {publisher}
  {{Springer}},\ \bibinfo {address} {{Berlin}},\ \bibinfo {year} {1981})\
  Chap.~\bibinfo {chapter} {21}, pp.\ \bibinfo {pages} {366--381}\BibitemShut
  {NoStop}%
\bibitem [{\citenamefont {Lizier}(2012)}]{Lizier2012Thesis}%
  \BibitemOpen
  \bibfield  {author} {\bibinfo {author} {\bibfnamefont {J.~T.}\ \bibnamefont
  {Lizier}},\ }\href@noop {} {{\selectlanguage {english}\emph {\bibinfo {title}
  {The local information dynamics of distributed computation in complex
  systems}}}},\ \bibinfo {edition} {2013th}\ ed.,\ Springer theses\ (\bibinfo
  {publisher} {Springer},\ \bibinfo {address} {Berlin, Germany},\ \bibinfo
  {year} {2012})\BibitemShut {NoStop}%
\bibitem [{\citenamefont {Edinburgh}\ \emph {et~al.}(2021)\citenamefont
  {Edinburgh}, \citenamefont {Eglen},\ and\ \citenamefont
  {Ercole}}]{Edinburgh2021Causality}%
  \BibitemOpen
  \bibfield  {author} {\bibinfo {author} {\bibfnamefont {T.}~\bibnamefont
  {Edinburgh}}, \bibinfo {author} {\bibfnamefont {S.~J.}\ \bibnamefont
  {Eglen}},\ and\ \bibinfo {author} {\bibfnamefont {A.}~\bibnamefont
  {Ercole}},\ }\bibfield  {title} {\bibinfo {title} {Causality indices for
  bivariate time series data: {{A}} comparative review of performance},\ }\href
  {https://doi.org/10.1063/5.0053519} {\bibfield  {journal} {\bibinfo
  {journal} {Chaos: An Interdisciplinary Journal of Nonlinear Science}\
  }\textbf {\bibinfo {volume} {31}},\ \bibinfo {pages} {083111} (\bibinfo
  {year} {2021})}\BibitemShut {NoStop}%
\bibitem [{\citenamefont {Bellman}(1966)}]{Bellman1966DynamicPrograming}%
  \BibitemOpen
  \bibfield  {author} {\bibinfo {author} {\bibfnamefont {R.}~\bibnamefont
  {Bellman}},\ }\bibfield  {title} {\bibinfo {title} {Dynamic programming},\
  }\href {https://doi.org/10.1126/science.153.3731.34} {\bibfield  {journal}
  {\bibinfo  {journal} {Science}\ }\textbf {\bibinfo {volume} {153}},\ \bibinfo
  {pages} {34} (\bibinfo {year} {1966})}\BibitemShut {NoStop}%
\bibitem [{\citenamefont {Donoho}(2000)}]{Donoho2000HighDimensionDataAnalysis}%
  \BibitemOpen
  \bibfield  {author} {\bibinfo {author} {\bibfnamefont {D.}~\bibnamefont
  {Donoho}},\ }\bibfield  {title} {\bibinfo {title} {High-dimensional data
  analysis: {{The}} curses and blessings of dimensionality},\ }in\ \href@noop
  {} {\emph {\bibinfo {booktitle} {AMS Math Challenges Lecture}}}\ (\bibinfo
  {year} {2000})\ pp.\ \bibinfo {pages} {1--32}\BibitemShut {NoStop}%
\bibitem [{\citenamefont {Staniek}\ and\ \citenamefont
  {Lehnertz}(2008)}]{Staniek2008SymbolicTE}%
  \BibitemOpen
  \bibfield  {author} {\bibinfo {author} {\bibfnamefont {M.}~\bibnamefont
  {Staniek}}\ and\ \bibinfo {author} {\bibfnamefont {K.}~\bibnamefont
  {Lehnertz}},\ }\bibfield  {title} {\bibinfo {title} {Symbolic transfer
  entropy},\ }\href {https://doi.org/10.1103/PhysRevLett.100.158101} {\bibfield
   {journal} {\bibinfo  {journal} {Physical Review Letters}\ }\textbf {\bibinfo
  {volume} {100}},\ \bibinfo {pages} {158101} (\bibinfo {year}
  {2008})}\BibitemShut {NoStop}%
\bibitem [{\citenamefont {Zanin}\ \emph {et~al.}(2012)\citenamefont {Zanin},
  \citenamefont {Zunino}, \citenamefont {Rosso},\ and\ \citenamefont
  {Papo}}]{Zanin2012PermunationEntropyReview}%
  \BibitemOpen
  \bibfield  {author} {\bibinfo {author} {\bibfnamefont {M.}~\bibnamefont
  {Zanin}}, \bibinfo {author} {\bibfnamefont {L.}~\bibnamefont {Zunino}},
  \bibinfo {author} {\bibfnamefont {O.~A.}\ \bibnamefont {Rosso}},\ and\
  \bibinfo {author} {\bibfnamefont {D.}~\bibnamefont {Papo}},\ }\bibfield
  {title} {\bibinfo {title} {Permutation entropy and its main biomedical and
  econophysics applications: {{A}} review},\ }\href
  {https://doi.org/10.3390/e14081553} {\bibfield  {journal} {\bibinfo
  {journal} {Entropy. An International and Interdisciplinary Journal of Entropy
  and Information Studies}\ }\textbf {\bibinfo {volume} {14}},\ \bibinfo
  {pages} {1553} (\bibinfo {year} {2012})}\BibitemShut {NoStop}%
\bibitem [{\citenamefont {Shorten}\ \emph {et~al.}(2021)\citenamefont
  {Shorten}, \citenamefont {Spinney},\ and\ \citenamefont
  {Lizier}}]{Shorten2021TESpiketrain}%
  \BibitemOpen
  \bibfield  {author} {\bibinfo {author} {\bibfnamefont {D.~P.}\ \bibnamefont
  {Shorten}}, \bibinfo {author} {\bibfnamefont {R.~E.}\ \bibnamefont
  {Spinney}},\ and\ \bibinfo {author} {\bibfnamefont {J.~T.}\ \bibnamefont
  {Lizier}},\ }\bibfield  {title} {\bibinfo {title} {Estimating transfer
  entropy in continuous time between neural spike trains or other event-based
  data},\ }\href {https://doi.org/10.1371/journal.pcbi.1008054} {\bibfield
  {journal} {\bibinfo  {journal} {PLOS Computational Biology}\ }\textbf
  {\bibinfo {volume} {17}},\ \bibinfo {pages} {1} (\bibinfo {year}
  {2021})}\BibitemShut {NoStop}%
\bibitem [{\citenamefont
  {Fulcher}(2018)}]{Fulcher2018:FeaturebasedTimeseriesAnalysis}%
  \BibitemOpen
  \bibfield  {author} {\bibinfo {author} {\bibfnamefont {B.~D.}\ \bibnamefont
  {Fulcher}},\ }\bibfield  {title} {\bibinfo {title} {Feature-based time-series
  analysis},\ }in\ \href {https://doi.org/10.1201/9781315181080-4} {\emph
  {\bibinfo {booktitle} {Feature Engineering for Machine Learning and Data
  Analytics}}}\ (\bibinfo  {publisher} {{CRC Press}},\ \bibinfo {year}
  {2018})\BibitemShut {NoStop}%
\bibitem [{\citenamefont {Henderson}\ and\ \citenamefont
  {Fulcher}(2021)}]{Trent2021FeatureSetEvaluation}%
  \BibitemOpen
  \bibfield  {author} {\bibinfo {author} {\bibfnamefont {T.}~\bibnamefont
  {Henderson}}\ and\ \bibinfo {author} {\bibfnamefont {B.~D.}\ \bibnamefont
  {Fulcher}},\ }\bibfield  {title} {\bibinfo {title} {An empirical evaluation
  of time-series feature sets},\ }in\ \href
  {https://doi.org/10.1109/ICDMW53433.2021.00134} {\emph {\bibinfo {booktitle}
  {2021 International Conference on Data Mining Workshops ({{ICDMW}})}}}\
  (\bibinfo {year} {2021})\ pp.\ \bibinfo {pages} {1032--1038}\BibitemShut
  {NoStop}%
\bibitem [{\citenamefont {Fulcher}\ \emph {et~al.}(2013)\citenamefont
  {Fulcher}, \citenamefont {Little},\ and\ \citenamefont
  {Jones}}]{Fulcher2013hctsa}%
  \BibitemOpen
  \bibfield  {author} {\bibinfo {author} {\bibfnamefont {B.~D.}\ \bibnamefont
  {Fulcher}}, \bibinfo {author} {\bibfnamefont {M.~A.}\ \bibnamefont
  {Little}},\ and\ \bibinfo {author} {\bibfnamefont {N.~S.}\ \bibnamefont
  {Jones}},\ }\bibfield  {title} {\bibinfo {title} {Highly comparative
  time-series analysis: The empirical structure of time series and their
  methods},\ }\href {https://doi.org/10.1098/rsif.2013.0048} {\bibfield
  {journal} {\bibinfo  {journal} {Journal of The Royal Society Interface}\
  }\textbf {\bibinfo {volume} {10}},\ \bibinfo {pages} {20130048} (\bibinfo
  {year} {2013})}\BibitemShut {NoStop}%
\bibitem [{\citenamefont {Fulcher}\ and\ \citenamefont
  {Jones}(2017)}]{Fulcher2017:HctsaComputationalFramework}%
  \BibitemOpen
  \bibfield  {author} {\bibinfo {author} {\bibfnamefont {B.~D.}\ \bibnamefont
  {Fulcher}}\ and\ \bibinfo {author} {\bibfnamefont {N.~S.}\ \bibnamefont
  {Jones}},\ }\bibfield  {title} {\bibinfo {title} {{\emph{Hctsa}}: {{A}}
  computational framework for automated time-series phenotyping using massive
  feature extraction},\ }\href {https://doi.org/10.1016/j.cels.2017.10.001}
  {\bibfield  {journal} {\bibinfo  {journal} {Cell Systems}\ }\textbf {\bibinfo
  {volume} {5}},\ \bibinfo {pages} {527} (\bibinfo {year} {2017})}\BibitemShut
  {NoStop}%
\bibitem [{\citenamefont {Lubba}\ \emph {et~al.}(2019)\citenamefont {Lubba},
  \citenamefont {Sethi}, \citenamefont {Knaute}, \citenamefont {Schultz},
  \citenamefont {Fulcher},\ and\ \citenamefont {Jones}}]{Lubba2019Catch22}%
  \BibitemOpen
  \bibfield  {author} {\bibinfo {author} {\bibfnamefont {C.~H.}\ \bibnamefont
  {Lubba}}, \bibinfo {author} {\bibfnamefont {S.}~\bibnamefont {Sethi}},
  \bibinfo {author} {\bibfnamefont {P.}~\bibnamefont {Knaute}}, \bibinfo
  {author} {\bibfnamefont {S.}~\bibnamefont {Schultz}}, \bibinfo {author}
  {\bibfnamefont {B.}~\bibnamefont {Fulcher}},\ and\ \bibinfo {author}
  {\bibfnamefont {N.}~\bibnamefont {Jones}},\ }\bibfield  {title} {\bibinfo
  {title} {Catch22: {{CAnonical Time-series CHaracteristics}}: {{Selected}}
  through highly comparative time-series analysis},\ }\href
  {https://doi.org/10.1007/s10618-019-00647-x} {\bibfield  {journal} {\bibinfo
  {journal} {Data Mining and Knowledge Discovery}\ }\textbf {\bibinfo {volume}
  {33}},\ \bibinfo {pages} {1821} (\bibinfo {year} {2019})}\BibitemShut
  {NoStop}%
\bibitem [{\citenamefont {Christ}\ \emph {et~al.}(2018)\citenamefont {Christ},
  \citenamefont {Braun}, \citenamefont {Neuffer},\ and\ \citenamefont
  {{Kempa-Liehr}}}]{tsfresh}%
  \BibitemOpen
  \bibfield  {author} {\bibinfo {author} {\bibfnamefont {M.}~\bibnamefont
  {Christ}}, \bibinfo {author} {\bibfnamefont {N.}~\bibnamefont {Braun}},
  \bibinfo {author} {\bibfnamefont {J.}~\bibnamefont {Neuffer}},\ and\ \bibinfo
  {author} {\bibfnamefont {A.~W.}\ \bibnamefont {{Kempa-Liehr}}},\ }\bibfield
  {title} {\bibinfo {title} {Time series {{FeatuRe}} extraction on basis of
  scalable hypothesis tests (tsfresh -- a python package)},\ }\href
  {https://doi.org/10.1016/j.neucom.2018.03.067} {\bibfield  {journal}
  {\bibinfo  {journal} {Neurocomputing}\ }\textbf {\bibinfo {volume} {307}},\
  \bibinfo {pages} {72} (\bibinfo {year} {2018})}\BibitemShut {NoStop}%
\bibitem [{\citenamefont {{O'Hara-Wild}}\ \emph {et~al.}(2021)\citenamefont
  {{O'Hara-Wild}}, \citenamefont {Hyndman},\ and\ \citenamefont
  {Wang}}]{feasts_pkg}%
  \BibitemOpen
  \bibfield  {author} {\bibinfo {author} {\bibfnamefont {M.}~\bibnamefont
  {{O'Hara-Wild}}}, \bibinfo {author} {\bibfnamefont {R.}~\bibnamefont
  {Hyndman}},\ and\ \bibinfo {author} {\bibfnamefont {E.}~\bibnamefont
  {Wang}},\ }\href@noop {} {\emph {\bibinfo {title} {Feasts: {{Feature}}
  Extraction and Statistics for Time Series}}} (\bibinfo {year}
  {2021})\BibitemShut {NoStop}%
\bibitem [{\citenamefont {Cover}\ and\ \citenamefont
  {Thomas}(2005)}]{CoverThomas2005InformationTheory}%
  \BibitemOpen
  \bibfield  {author} {\bibinfo {author} {\bibfnamefont {T.}~\bibnamefont
  {Cover}}\ and\ \bibinfo {author} {\bibfnamefont {J.}~\bibnamefont {Thomas}},\
  }\bibfield  {title} {\bibinfo {title} {Information theory and statistics},\
  }in\ \href {https://doi.org/10.1002/047174882X.ch11} {\emph {\bibinfo
  {booktitle} {Elements of Information Theory}}}\ (\bibinfo  {publisher} {{John
  Wiley \& Sons, Ltd}},\ \bibinfo {year} {2005})\ Chap.~\bibinfo {chapter}
  {11}, pp.\ \bibinfo {pages} {347--408}\BibitemShut {NoStop}%
\bibitem [{\citenamefont {Lizier}(2014)}]{Lizier2014JIDT}%
  \BibitemOpen
  \bibfield  {author} {\bibinfo {author} {\bibfnamefont {J.~T.}\ \bibnamefont
  {Lizier}},\ }\bibfield  {title} {\bibinfo {title} {Jidt: An
  information-theoretic toolkit for studying the dynamics of complex systems},\
  }\bibfield  {journal} {\bibinfo  {journal} {Frontiers in Robotics and AI}\
  }\textbf {\bibinfo {volume} {1}},\ \href
  {https://doi.org/10.3389/frobt.2014.00011} {10.3389/frobt.2014.00011}
  (\bibinfo {year} {2014})\BibitemShut {NoStop}%
\bibitem [{\citenamefont {Verdes}(2005)}]{Verdes2005CausalityAssessment}%
  \BibitemOpen
  \bibfield  {author} {\bibinfo {author} {\bibfnamefont {P.~F.}\ \bibnamefont
  {Verdes}},\ }\bibfield  {title} {\bibinfo {title} {Assessing causality from
  multivariate time series},\ }\href
  {https://doi.org/10.1103/PhysRevE.72.026222} {\bibfield  {journal} {\bibinfo
  {journal} {Physical Review E}\ }\textbf {\bibinfo {volume} {72}},\ \bibinfo
  {pages} {026222} (\bibinfo {year} {2005})}\BibitemShut {NoStop}%
\bibitem [{\citenamefont {Vicente}\ \emph {et~al.}(2011)\citenamefont
  {Vicente}, \citenamefont {Wibral}, \citenamefont {Lindner},\ and\
  \citenamefont {Pipa}}]{Vicente2011TE}%
  \BibitemOpen
  \bibfield  {author} {\bibinfo {author} {\bibfnamefont {R.}~\bibnamefont
  {Vicente}}, \bibinfo {author} {\bibfnamefont {M.}~\bibnamefont {Wibral}},
  \bibinfo {author} {\bibfnamefont {M.}~\bibnamefont {Lindner}},\ and\ \bibinfo
  {author} {\bibfnamefont {G.}~\bibnamefont {Pipa}},\ }\bibfield  {title}
  {\bibinfo {title} {Transfer entropy---a model-free measure of effective
  connectivity for the neurosciences},\ }\href
  {https://doi.org/10.1007/s10827-010-0262-3} {\bibfield  {journal} {\bibinfo
  {journal} {Journal of Computational Neuroscience}\ }\textbf {\bibinfo
  {volume} {30}},\ \bibinfo {pages} {45} (\bibinfo {year} {2011})}\BibitemShut
  {NoStop}%
\bibitem [{\citenamefont {Holm}(1979)}]{Holm1979MultipleTest}%
  \BibitemOpen
  \bibfield  {author} {\bibinfo {author} {\bibfnamefont {S.}~\bibnamefont
  {Holm}},\ }\bibfield  {title} {\bibinfo {title} {A simple sequentially
  rejective multiple test procedure},\ }\href@noop {} {\bibfield  {journal}
  {\bibinfo  {journal} {Scandinavian Journal of Statistics}\ }\textbf {\bibinfo
  {volume} {6}},\ \bibinfo {pages} {65} (\bibinfo {year} {1979})}\BibitemShut
  {NoStop}%
\bibitem [{\citenamefont {Kraskov}\ \emph {et~al.}(2004)\citenamefont
  {Kraskov}, \citenamefont {St{\"o}gbauer},\ and\ \citenamefont
  {Grassberger}}]{Kraskov2004MI}%
  \BibitemOpen
  \bibfield  {author} {\bibinfo {author} {\bibfnamefont {A.}~\bibnamefont
  {Kraskov}}, \bibinfo {author} {\bibfnamefont {H.}~\bibnamefont
  {St{\"o}gbauer}},\ and\ \bibinfo {author} {\bibfnamefont {P.}~\bibnamefont
  {Grassberger}},\ }\bibfield  {title} {\bibinfo {title} {Estimating mutual
  information},\ }\href {https://doi.org/10.1103/PhysRevE.69.066138} {\bibfield
   {journal} {\bibinfo  {journal} {Physical Review E}\ }\textbf {\bibinfo
  {volume} {69}},\ \bibinfo {pages} {066138} (\bibinfo {year}
  {2004})}\BibitemShut {NoStop}%
\bibitem [{\citenamefont {Cliff}\ \emph {et~al.}(2021)\citenamefont {Cliff},
  \citenamefont {Novelli}, \citenamefont {Fulcher}, \citenamefont {Shine},\
  and\ \citenamefont {Lizier}}]{Cliff2021SignificanceTesting}%
  \BibitemOpen
  \bibfield  {author} {\bibinfo {author} {\bibfnamefont {O.~M.}\ \bibnamefont
  {Cliff}}, \bibinfo {author} {\bibfnamefont {L.}~\bibnamefont {Novelli}},
  \bibinfo {author} {\bibfnamefont {B.~D.}\ \bibnamefont {Fulcher}}, \bibinfo
  {author} {\bibfnamefont {J.~M.}\ \bibnamefont {Shine}},\ and\ \bibinfo
  {author} {\bibfnamefont {J.~T.}\ \bibnamefont {Lizier}},\ }\bibfield  {title}
  {\bibinfo {title} {Assessing the significance of directed and multivariate
  measures of linear dependence between time series},\ }\href
  {https://doi.org/10.1103/PhysRevResearch.3.013145} {\bibfield  {journal}
  {\bibinfo  {journal} {Physical Review Research}\ }\textbf {\bibinfo {volume}
  {3}},\ \bibinfo {pages} {013145} (\bibinfo {year} {2021})}\BibitemShut
  {NoStop}%
\bibitem [{\citenamefont {Goswami}(2019)}]{Goswami2019NonlinearTS}%
  \BibitemOpen
  \bibfield  {author} {\bibinfo {author} {\bibfnamefont {B.}~\bibnamefont
  {Goswami}},\ }\bibfield  {title} {\bibinfo {title} {A brief introduction to
  nonlinear time series analysis and recurrence plots},\ }\href
  {https://doi.org/10.3390/vibration2040021} {\bibfield  {journal} {\bibinfo
  {journal} {Vibration}\ }\textbf {\bibinfo {volume} {2}},\ \bibinfo {pages}
  {332} (\bibinfo {year} {2019})}\BibitemShut {NoStop}%
\bibitem [{\citenamefont {Schreiber}\ and\ \citenamefont
  {Schmitz}(2000)}]{Schreiber2000Surrogates}%
  \BibitemOpen
  \bibfield  {author} {\bibinfo {author} {\bibfnamefont {T.}~\bibnamefont
  {Schreiber}}\ and\ \bibinfo {author} {\bibfnamefont {A.}~\bibnamefont
  {Schmitz}},\ }\bibfield  {title} {\bibinfo {title} {Surrogate time series},\
  }\href {https://doi.org/10.1016/S0167-2789(00)00043-9} {\bibfield  {journal}
  {\bibinfo  {journal} {Physica D: Nonlinear Phenomena}\ }\textbf {\bibinfo
  {volume} {142}},\ \bibinfo {pages} {346} (\bibinfo {year}
  {2000})}\BibitemShut {NoStop}%
\bibitem [{\citenamefont {Vicente}\ and\ \citenamefont
  {Wibral}(2014)}]{Vicente2014InformationTransfer}%
  \BibitemOpen
  \bibfield  {author} {\bibinfo {author} {\bibfnamefont {R.}~\bibnamefont
  {Vicente}}\ and\ \bibinfo {author} {\bibfnamefont {M.}~\bibnamefont
  {Wibral}},\ }\bibfield  {title} {\bibinfo {title} {Efficient estimation of
  information transfer},\ }in\ \href
  {https://doi.org/10.1007/978-3-642-54474-3_2} {\emph {\bibinfo {booktitle}
  {Directed Information Measures in Neuroscience}}},\ \bibinfo {editor} {edited
  by\ \bibinfo {editor} {\bibfnamefont {M.}~\bibnamefont {Wibral}}, \bibinfo
  {editor} {\bibfnamefont {R.}~\bibnamefont {Vicente}},\ and\ \bibinfo {editor}
  {\bibfnamefont {J.~T.}\ \bibnamefont {Lizier}}}\ (\bibinfo  {publisher}
  {{Springer Berlin Heidelberg}},\ \bibinfo {address} {{Berlin, Heidelberg}},\
  \bibinfo {year} {2014})\ pp.\ \bibinfo {pages} {37--58}\BibitemShut {NoStop}%
\bibitem [{\citenamefont {Wibral}\ \emph {et~al.}(2014)\citenamefont {Wibral},
  \citenamefont {Vicente},\ and\ \citenamefont
  {Lindner}}]{Wibral2014TEinNeuroscience}%
  \BibitemOpen
  \bibfield  {author} {\bibinfo {author} {\bibfnamefont {M.}~\bibnamefont
  {Wibral}}, \bibinfo {author} {\bibfnamefont {R.}~\bibnamefont {Vicente}},\
  and\ \bibinfo {author} {\bibfnamefont {M.}~\bibnamefont {Lindner}},\
  }\bibfield  {title} {\bibinfo {title} {Transfer entropy in neuroscience},\
  }in\ \href {https://doi.org/10.1007/978-3-642-54474-3_1} {\emph {\bibinfo
  {booktitle} {Directed Information Measures in Neuroscience}}},\ \bibinfo
  {editor} {edited by\ \bibinfo {editor} {\bibfnamefont {M.}~\bibnamefont
  {Wibral}}, \bibinfo {editor} {\bibfnamefont {R.}~\bibnamefont {Vicente}},\
  and\ \bibinfo {editor} {\bibfnamefont {J.~T.}\ \bibnamefont {Lizier}}}\
  (\bibinfo  {publisher} {{Springer Berlin Heidelberg}},\ \bibinfo {address}
  {{Berlin, Heidelberg}},\ \bibinfo {year} {2014})\ pp.\ \bibinfo {pages}
  {3--36}\BibitemShut {NoStop}%
\bibitem [{\citenamefont {S.~J.~Leybourne}\ and\ \citenamefont
  {Tremayne}(1996)}]{Leybourne1996NonstationarityTSEconomics}%
  \BibitemOpen
  \bibfield  {author} {\bibinfo {author} {\bibfnamefont {B.~P. M.~M.}\
  \bibnamefont {S.~J.~Leybourne}}\ and\ \bibinfo {author} {\bibfnamefont
  {A.~R.}\ \bibnamefont {Tremayne}},\ }\bibfield  {title} {\bibinfo {title}
  {Can economic time series be differenced to stationarity?},\ }\href
  {https://doi.org/10.1080/07350015.1996.10524673} {\bibfield  {journal}
  {\bibinfo  {journal} {Journal of Business \& Economic Statistics}\ }\textbf
  {\bibinfo {volume} {14}},\ \bibinfo {pages} {435} (\bibinfo {year}
  {1996})}\BibitemShut {NoStop}%
\bibitem [{\citenamefont {Cassidy}\ and\ \citenamefont
  {Penny}(2002)}]{Cassidy2002BayesianARModelBiomed}%
  \BibitemOpen
  \bibfield  {author} {\bibinfo {author} {\bibfnamefont {M.}~\bibnamefont
  {Cassidy}}\ and\ \bibinfo {author} {\bibfnamefont {W.}~\bibnamefont
  {Penny}},\ }\bibfield  {title} {\bibinfo {title} {Bayesian nonstationary
  autoregressive models for biomedical signal analysis},\ }\href
  {https://doi.org/10.1109/TBME.2002.803511} {\bibfield  {journal} {\bibinfo
  {journal} {IEEE Transactions on Biomedical Engineering}\ }\textbf {\bibinfo
  {volume} {49}},\ \bibinfo {pages} {1142} (\bibinfo {year}
  {2002})}\BibitemShut {NoStop}%
\bibitem [{\citenamefont {Box}\ and\ \citenamefont
  {Jenkins}(1976)}]{BoxJenkins76}%
  \BibitemOpen
  \bibfield  {author} {\bibinfo {author} {\bibfnamefont {{\relax
  George.E.P}.}~\bibnamefont {Box}}\ and\ \bibinfo {author} {\bibfnamefont
  {G.~M.}\ \bibnamefont {Jenkins}},\ }\href@noop {} {\emph {\bibinfo {title}
  {Time Series Analysis: {{Forecasting}} and Control}}}\ (\bibinfo  {publisher}
  {{Holden-Day}},\ \bibinfo {year} {1976})\ Chap.~\bibinfo {chapter}
  {3}\BibitemShut {NoStop}%
\bibitem [{\citenamefont {Venables}\ and\ \citenamefont
  {Ripley}(2002)}]{Veneables2002ModernStats}%
  \BibitemOpen
  \bibfield  {author} {\bibinfo {author} {\bibfnamefont {B.}~\bibnamefont
  {Venables}}\ and\ \bibinfo {author} {\bibfnamefont {B.}~\bibnamefont
  {Ripley}},\ }\href {https://doi.org/10.1007/b97626} {\emph {\bibinfo {title}
  {Modern Applied Statistics with {{S}}}}}\ (\bibinfo  {publisher}
  {{Springer}},\ \bibinfo {year} {2002})\BibitemShut {NoStop}%
\bibitem [{\citenamefont {Feldman}(2012)}]{Feldman2012STDP}%
  \BibitemOpen
  \bibfield  {author} {\bibinfo {author} {\bibfnamefont {D.}~\bibnamefont
  {Feldman}},\ }\bibfield  {title} {\bibinfo {title} {The spike-timing
  dependence of plasticity},\ }\href
  {https://doi.org/https://doi.org/10.1016/j.neuron.2012.08.001} {\bibfield
  {journal} {\bibinfo  {journal} {Neuron}\ }\textbf {\bibinfo {volume} {75}},\
  \bibinfo {pages} {556} (\bibinfo {year} {2012})}\BibitemShut {NoStop}%
\bibitem [{\citenamefont {Usrey}\ \emph {et~al.}(2000)\citenamefont {Usrey},
  \citenamefont {Alonso},\ and\ \citenamefont
  {Reid}}]{Usrey2000SynapticInteractions}%
  \BibitemOpen
  \bibfield  {author} {\bibinfo {author} {\bibfnamefont {W.~M.}\ \bibnamefont
  {Usrey}}, \bibinfo {author} {\bibfnamefont {J.-M.}\ \bibnamefont {Alonso}},\
  and\ \bibinfo {author} {\bibfnamefont {R.~C.}\ \bibnamefont {Reid}},\
  }\bibfield  {title} {\bibinfo {title} {Synaptic interactions between thalamic
  inputs to simple cells in cat visual cortex},\ }\href
  {https://doi.org/10.1523/JNEUROSCI.20-14-05461.2000} {\bibfield  {journal}
  {\bibinfo  {journal} {Journal of Neuroscience}\ }\textbf {\bibinfo {volume}
  {20}},\ \bibinfo {pages} {5461} (\bibinfo {year} {2000})}\BibitemShut
  {NoStop}%
\bibitem [{\citenamefont {Gerstner}\ \emph {et~al.}(1997)\citenamefont
  {Gerstner}, \citenamefont {Kreiter}, \citenamefont {Markram},\ and\
  \citenamefont {{Andreas V. M. Herz}}}]{Gerstner1997NeuralCodes}%
  \BibitemOpen
  \bibfield  {author} {\bibinfo {author} {\bibfnamefont {W.}~\bibnamefont
  {Gerstner}}, \bibinfo {author} {\bibfnamefont {A.~K.}\ \bibnamefont
  {Kreiter}}, \bibinfo {author} {\bibfnamefont {H.}~\bibnamefont {Markram}},\
  and\ \bibinfo {author} {\bibnamefont {{Andreas V. M. Herz}}},\ }\bibfield
  {title} {\bibinfo {title} {Neural codes: {{Firing}} rates and beyond},\
  }\href {https://doi.org/10.1073/pnas.94.24.12740} {\bibfield  {journal}
  {\bibinfo  {journal} {Proceedings of the National Academy of Sciences}\
  }\textbf {\bibinfo {volume} {94}},\ \bibinfo {pages} {12740} (\bibinfo {year}
  {1997})}\BibitemShut {NoStop}%
\bibitem [{\citenamefont
  {Schreiber}(1997)}]{Schreiber1997DetectingNonstationarity}%
  \BibitemOpen
  \bibfield  {author} {\bibinfo {author} {\bibfnamefont {T.}~\bibnamefont
  {Schreiber}},\ }\bibfield  {title} {\bibinfo {title} {Detecting and analyzing
  nonstationarity in a time series using nonlinear cross predictions},\ }\href
  {https://doi.org/10.1103/PhysRevLett.78.843} {\bibfield  {journal} {\bibinfo
  {journal} {Physical Review Letters}\ }\textbf {\bibinfo {volume} {78}},\
  \bibinfo {pages} {843} (\bibinfo {year} {1997})}\BibitemShut {NoStop}%
\bibitem [{\citenamefont {Aguirre}\ and\ \citenamefont
  {Letellier}(2012)}]{Aguirre2012Nonstationarity}%
  \BibitemOpen
  \bibfield  {author} {\bibinfo {author} {\bibfnamefont {L.~A.}\ \bibnamefont
  {Aguirre}}\ and\ \bibinfo {author} {\bibfnamefont {C.}~\bibnamefont
  {Letellier}},\ }\bibfield  {title} {\bibinfo {title} {Nonstationarity
  signatures in the dynamics of global nonlinear models},\ }\href
  {https://doi.org/10.1063/1.4748852} {\bibfield  {journal} {\bibinfo
  {journal} {Chaos: An Interdisciplinary Journal of Nonlinear Science}\
  }\textbf {\bibinfo {volume} {22}},\ \bibinfo {pages} {033136} (\bibinfo
  {year} {2012})}\BibitemShut {NoStop}%
\bibitem [{\citenamefont {Griffith}\ and\ \citenamefont
  {Koch}(2014)}]{Griffith2014InfoSynergy}%
  \BibitemOpen
  \bibfield  {author} {\bibinfo {author} {\bibfnamefont {V.}~\bibnamefont
  {Griffith}}\ and\ \bibinfo {author} {\bibfnamefont {C.}~\bibnamefont
  {Koch}},\ }\bibfield  {title} {\bibinfo {title} {Quantifying synergistic
  mutual information},\ }in\ \href@noop {} {\emph {\bibinfo {booktitle} {Guided
  {{Self-Organization}}: {{Inception}}}}}\ (\bibinfo  {publisher} {{Springer
  Berlin Heidelberg}},\ \bibinfo {address} {{Berlin, Heidelberg}},\ \bibinfo
  {year} {2014})\ pp.\ \bibinfo {pages} {159--190}\BibitemShut {NoStop}%
\bibitem [{\citenamefont {Timme}\ \emph {et~al.}(2014)\citenamefont {Timme},
  \citenamefont {Alford}, \citenamefont {Flecker},\ and\ \citenamefont
  {Beggs}}]{Timme2014SynergyRedundancy}%
  \BibitemOpen
  \bibfield  {author} {\bibinfo {author} {\bibfnamefont {N.}~\bibnamefont
  {Timme}}, \bibinfo {author} {\bibfnamefont {W.}~\bibnamefont {Alford}},
  \bibinfo {author} {\bibfnamefont {B.}~\bibnamefont {Flecker}},\ and\ \bibinfo
  {author} {\bibfnamefont {J.~M.}\ \bibnamefont {Beggs}},\ }\bibfield  {title}
  {\bibinfo {title} {Synergy, redundancy, and multivariate information
  measures: an experimentalist's perspective},\ }\href
  {https://doi.org/10.1007/s10827-013-0458-4} {\bibfield  {journal} {\bibinfo
  {journal} {Journal of Computational Neuroscience}\ }\textbf {\bibinfo
  {volume} {36}},\ \bibinfo {pages} {119} (\bibinfo {year} {2014})}\BibitemShut
  {NoStop}%
\bibitem [{\citenamefont {Quax}\ \emph {et~al.}(2017)\citenamefont {Quax},
  \citenamefont {Har-Shemesh},\ and\ \citenamefont
  {Sloot}}]{Quax2017QuantifySynergisticInfo}%
  \BibitemOpen
  \bibfield  {author} {\bibinfo {author} {\bibfnamefont {R.}~\bibnamefont
  {Quax}}, \bibinfo {author} {\bibfnamefont {O.}~\bibnamefont {Har-Shemesh}},\
  and\ \bibinfo {author} {\bibfnamefont {P.~M.~A.}\ \bibnamefont {Sloot}},\
  }\bibfield  {title} {\bibinfo {title} {Quantifying synergistic information
  using intermediate stochastic variables},\ }\bibfield  {journal} {\bibinfo
  {journal} {Entropy}\ }\textbf {\bibinfo {volume} {19}},\ \href
  {https://doi.org/10.3390/e19020085} {10.3390/e19020085} (\bibinfo {year}
  {2017})\BibitemShut {NoStop}%
\bibitem [{\citenamefont {Rozell}\ and\ \citenamefont
  {Johnson}(2006)}]{Rozell2006RedundantPopulationCode}%
  \BibitemOpen
  \bibfield  {author} {\bibinfo {author} {\bibfnamefont {C.~J.}\ \bibnamefont
  {Rozell}}\ and\ \bibinfo {author} {\bibfnamefont {D.~H.}\ \bibnamefont
  {Johnson}},\ }\bibfield  {title} {\bibinfo {title} {Analyzing the robustness
  of redundant population codes in sensory and feature extraction systems},\
  }\href {https://doi.org/https://doi.org/10.1016/j.neucom.2005.12.079}
  {\bibfield  {journal} {\bibinfo  {journal} {Neurocomputing}\ }\textbf
  {\bibinfo {volume} {69}},\ \bibinfo {pages} {1215} (\bibinfo {year}
  {2006})},\ \bibinfo {note} {computational Neuroscience: Trends in Research
  2006}\BibitemShut {NoStop}%
\bibitem [{\citenamefont {Rozell}\ and\ \citenamefont
  {Johnson}(2005)}]{Rozell2005RedundancyNoiseReduction}%
  \BibitemOpen
  \bibfield  {author} {\bibinfo {author} {\bibfnamefont {C.}~\bibnamefont
  {Rozell}}\ and\ \bibinfo {author} {\bibfnamefont {D.}~\bibnamefont
  {Johnson}},\ }\bibfield  {title} {\bibinfo {title} {Analysis of noise
  reduction in redundant expansions under distributed processing
  requirements},\ }in\ \href {https://doi.org/10.1109/ICASSP.2005.1415976}
  {\emph {\bibinfo {booktitle} {Proceedings. (ICASSP '05). IEEE International
  Conference on Acoustics, Speech, and Signal Processing, 2005.}}},\
  Vol.~\bibinfo {volume} {4}\ (\bibinfo {year} {2005})\ pp.\ \bibinfo {pages}
  {iv/185--iv/188 Vol. 4}\BibitemShut {NoStop}%
\bibitem [{\citenamefont {Tishby}\ \emph {et~al.}(1999)\citenamefont {Tishby},
  \citenamefont {Pereira},\ and\ \citenamefont
  {Bialek}}]{Tishby1999InformationBottleneck}%
  \BibitemOpen
  \bibfield  {author} {\bibinfo {author} {\bibfnamefont {N.}~\bibnamefont
  {Tishby}}, \bibinfo {author} {\bibfnamefont {F.~C.}\ \bibnamefont
  {Pereira}},\ and\ \bibinfo {author} {\bibfnamefont {W.}~\bibnamefont
  {Bialek}},\ }\bibfield  {title} {\bibinfo {title} {The information bottleneck
  method},\ }in\ \href {https://arxiv.org/abs/physics/0004057} {\emph {\bibinfo
  {booktitle} {Proc. of the 37-th Annual Allerton Conference on Communication,
  Control and Computing}}}\ (\bibinfo {year} {1999})\ pp.\ \bibinfo {pages}
  {368--377}\BibitemShut {NoStop}%
\end{thebibliography}%

\clearpage

\onecolumngrid
\appendix
\section{}

\begin{table}[h]
\caption{\label{tab:FeatureOverview} \textit{Catch22} Feature Overview. 
This feature set comprises 22 features that cover a broad range of time-series properties, including the distribution of values in the time series, linear and nonlinear temporal autocorrelation properties, and others \cite{Lubba2019Catch22}. 
The feature names are adopted from the larger \textbf{hctsa} feature set \cite{Fulcher2013hctsa}. 
This table presents the feature short names used throughout the paper and their corresponding descriptions.}
\begin{ruledtabular}
\begin{tabular}{|p{7.2cm}|p{3cm}|p{7.2cm}|}
    \textbf{Feature Name} & \textbf{Short Name} & \textbf{Description} \\ \hline
    \verb|DN_HistogramMode_5| & \verb|mode5| & 5-bin histogram mode \\
    \verb|DN_HistogramMode_10| & \verb|mode10| & 10-bin histogram mode \\
    \verb|DN_OutlierInclude_p_001_mdrmd| & \verb|outlier_timing_pos| & Positive outlier timing \\
    \verb|DN_OutlierInclude_n_001_mdrmd| & \verb|outlier_timing_neg| & Negative outlier timing \\
    \verb|ﬁrst1e_acf_tau| & \verb|acf_timescale| & First $1/e$ crossing of the ACF \\
    \verb|ﬁrstMin_acf| & \verb|acf_first_min| & First minimum of the ACF \\
    \verb|SP_Summaries_welch_rect_area_5_1| & \verb|low_freq_power| & Power in lowest 20\% frequencies \\
    \verb|SP_Summaries_welch_rect_centroid| & \verb|centroid_freq| & Centroid frequency \\
    \verb|FC_LocalSimple_mean3_stderr| & \verb|forecast_error| & Error of 3-point rolling mean forecast \\
    \verb|FC_LocalSimple_mean1_tauresrat| & \verb|whiten_timescale| & Change in autocorrelation timescale after incremental differencing \\
    \verb|MD_hrv_classic_pnn40| & \verb|high_fluctuation| & Proportion of high incremental changes in the series \\  
    \verb|SB_BinaryStats_mean_longstretch1| & \verb|stretch_high| & Longest stretch of above-mean values \\
    \verb|SB_BinaryStats_diff_longstretch0| & \verb|stretch_decreasing| & Longest stretch of decreasing values \\
    \verb|SB_MotifThree_quantile_hh| & \verb|entropy_pairs| & Entropy of successive pairs in symbolized series \\
    \verb|CO_HistogramAMI_even_2_5| & \verb|ami2| & Histogram-based automutual information (lag 2, 5 bins) \\
    \verb|CO_trev_1_num| & \verb|trev| & Time reversibility \\
    \verb|IN_AutoMutualInfoStats_40_gaussian_fmmi| & \verb|ami_timescale| & First minimum of the AMI function \\
    \verb|SB_TransitionMatrix_3ac_sumdiagcov| & \verb|transition_variance| & Transition matrix column variance \\
    \verb|PD_PeriodicityWang_th001| & \verb|periodicity| & Wang's periodicity metric \\
    \verb|CO_Embed2_Dist_tau_d_expfit_meandiff| & \verb|embedding_dist| & Goodness of exponential fit to embedding distance distribution \\
    \verb|SC_FluctAnal_2_rsrangeﬁt_50_1_logi_prop_r1| & \verb|rs_range| & Rescaled range fluctuation analysis (low-scale scaling) \\
    \verb|SC_FluctAnal_2_dfa_50_1_2_logi_prop_r1| & \verb|dfa| & Detrended fluctuation analysis (low-scale scaling) \\
\end{tabular}
\end{ruledtabular}
\end{table}

\end{document}